\documentclass[showkeys,%
preprint%
amsmath,amssymb,%
aps,%
pre,%
groupedaddress,%
superscriptaddress,%
a4paper,%
floatfix,%
eprint,%
preprintnumbers,%
eqsecnum,%
nofootinbib,%
]{revtex4-2}

\usepackage[utf8]{inputenc}
\usepackage{fontenc}
\usepackage{amsmath}

\usepackage{subfigure}
\usepackage{multirow}
\usepackage{physics}
\usepackage{dcolumn}

\usepackage{bm}
\usepackage{graphics,graphicx} 
\usepackage{natbib}
\usepackage{xcolor,color}
\usepackage{hyperref}
\hypersetup{colorlinks=true,
	breaklinks,
	hyperfootnotes,
	linkcolor={red},
	final,
	urlcolor={blue},
	citecolor=cyan,
	breaklinks=true}

\usepackage{booktabs}
\usepackage{siunitx}
\usepackage{float}
\usepackage{verbatim}
\usepackage{orcidlink}

\begin{document}
\normalsize
\title{Emergence of Multi-Scroll Attractors }


\author{Tanmayee Patra\,\orcidlink{0009-0007-5555-0068}}
\email{tanmayee\_patra@nitrkl.ac.in}
\affiliation{%
Department of Physics and Astronomy, National Institute of Technology Rourkela, Odisha-769008, India}%
\author{Biplab Ganguli\,\orcidlink{0000-0003-2583-5752}}%
\email{biplabg@nitrkl.ac.in}
\affiliation{%
Department of Physics and Astronomy, National Institute of Technology Rourkela, Odisha-769008, India}%

\begin{abstract}
Phase space trajectories are fundamentally important for understanding and analysing chaotic attractors. This is mostly carried out by direct numerical solution of the dynamical equations. Though the origin of scrolls can be understood from the properties of dynamical equations, their appearance in the phase space can also be inferred from the geometry and relative orientations of Nambu surfaces, drawn using Nambu Hamiltonians than from direct numerical solutions. Therefore, one can attribute the origin of wings in the phase space due to energy like Nambu surfaces, giving a geometrical interpretation. In this article, we have carried out, both numerical analysis of  bifurcation diagram and Lyapunov exponents(LEs) to characterise chaos and geometric approach by applying the Nambu generalized Hamiltonian mechanics to explain the fundamental reason for the appearance of wings like geometry in the phase space. We have shown how a fixed number of scrolls or wings can appear in the phase space due to specific geometry of the Nambu surfaces and how different geometries are formed when set of parameters are changed. 

\end{abstract}

\maketitle
\pagestyle{plain}
\section{\label{secI}Introduction}

The importance of phase space trajectories for understanding chaotic attractors cannot be overstated—they are the core framework for analysing, visualizing, and quantifying chaos. A breakdown of their critical roles are given here.  As we know, chaotic systems are deterministic but highly sensitive to initial conditions, making their long-term behaviour unpredictable. Henceforth, phase space trajectories are essential because they :
(a) Reveal the ``shape" of chaos (attractor geometry) which provide a geometric representation of the system's evolution, (b) Allow quantification of chaotic measures (Lyapunov exponents, fractal dimensions), (c) Distinguish chaos from noise or periodicity. Without phase space analysis, chaotic behaviour would remain a mathematical abstraction rather than a observable phenomenon. In other wards, we can say that it would be nearly impossible to ``see" chaos in action without phase space plots.

The terminology used for various attractors is directly influenced by the distinctive shapes of their chaotic attractors within the phase space. This connection underscores the significance of understanding these geometric features, as they provide valuable insights into their complex dynamics. Several number of literatures are listed here and they  indicate the fact that the realm of chaotic attractors is rich with diverse topological structures, such as multi-wing, multi-scroll, multi-star, multi-flower, multi-cavity, multi-torus, multi-layer spiral, multi-grid, multi-folded chaotic attractors \cite{SAHOO2022927, wang2019novel,lin2022design,wang2007simulation,yang2003twin,xiao2020constructing,signing2018coexistence,SAHOO2022112598,du2021multiring,DING20241053,Yan2022,zidan2012controllable,yu2006general,zhang2018one,goufo2022linear,shen2025extremely}.

Henceforth, The scroll(s) or wing(s) in chaotic attractors presents an intriguing and complex structure with lobe(s) or butterfly shape like structures visible in their phase portraits. This feature is intricately tied to the topology of the attractor, showcasing a remarkable range that can extend from just two scrolls/wings to an impressive multitude. The study of these attractors not only deepens our understanding of dynamical systems but also opens exciting avenues for exploration in various scientific fields. A simple definition of a scroll/wing is also mentioned by Guoyuan Qi et al. in their publications\cite{wang20093,qi2008four}.

The creation of complex multi-scroll and multi-wings chaotic attractors from 3-D autonomous systems has progressed rapidly since last three decades. Multi-wing and multi-scroll attractors are  types of chaotic attractors that exhibit complex dynamical behaviour. They are generated around multiple equilibrium points. A multi-wing attractor is a chaotic system whose phase portrait resembles multiple ``wings" (or lobes) symmetrically or asymmetrically arranged around a central point, where wings(popularly known as butterfly effect) are formed by non-linear folding and stretching of the trajectories. It typically has a single equilibrium point (or a few), but the trajectories swirl around in a way that creates multiple distinct wing-like structures. In contrast, a multi-scroll attractor consists of multiple ``scrolls" (or chaotic saddle foci) arranged in a grid-like (1D, 2D, or 3D) structure and it has multiple equilibrium points, each contributing to a separate scroll. The trajectories cycle around these points, creating a more segmented appearance compared to multi-wing attractors. In this article, our main objective is to understand how number of wings of an attractor are formed out of energy like Nambu surfaces in the phase space. 

A multi-scroll attrator is a system which is an extension of the classic Chua's circuit and other non-linear oscillators, designed to produce more intricate chaotic behaviour with multiple equilibrium points. The key features of multi-scroll attractors are:(i)they have multiple equilibrium points and each ``scroll" corresponds to a region around an unstable equilibrium point,(ii)they are generated via piecewise-linear (PWL) functions (e.g. stair-step functions) \cite{CHEN201622, belykh2023hidden, 7937684} or smooth nonlinearities (e.g. hyperbolic tangent functions) \cite{leutcho2018dynamical, SIGNING2018263, LI2008387} ,(iii)they can have controlled chaos means the number of scrolls can be programmed by adjusting system parameters \cite{CHENG2023113837, Fan2024, XIONG20182381, ZHANG2024115109, ZHANG2018793}. Suykens first constructed a multi-scroll system based on the Chua circuit \cite{PhysRevA.30.1155}. Later on, construction of various multi-scroll attractor based on Chen system took place  \cite{zhong2002systematic, lu2006brief, lu2006generating}. Due to all these features multi-scroll system has various applications such as: (i)for secure communication\cite{ZHANG2024115109, Fan2024,Wang2009}, because multi-scroll chaos provides higher entropy, making encryption harder to crack, (ii)for random number generation in cryptography \cite{adelakun2023robust, lin2022generating},(iii)in chaotic neural networks for associative memory and optimization\cite{CHENG2023113837, SILVAJUAREZ2021125831, DOUNGMOGOUFO2022112283}, (iv)in biological systems modelling\cite{WEBER2018457} etc.

Furthermore, a multi-scroll system can generate  single-, double-, and multi-scroll patterns even infinite scroll with infinite equlibrium points\cite{KINGNI2017209} also. It is also capable of self-reproducing these dynamic characteristics by controlling the system's initial conditions \cite{HU2025826, li2023multi, He_2023} . In the paper \cite{leutcho2018dynamical,vaidyanathan2023novel}, the authors demonstrated that the Jerk circuit, which incorporates the sine function, can generate multi-scroll attractors, with the number of scrolls being influenced by the simulation time. Multi-scroll chaotic systems have a more complex dynamical structure and richer behaviours than single-scroll and double-scroll systems. Understanding the generation mechanisms of multi-scroll attractors is essential for advancing chaos theory. 

A range of numerical diagnostic tools—such as bifurcation diagrams, Lyapunov exponents, Poincaré maps, power spectra, and fractal dimensions—are essential for understanding the complex dynamics of chaotic attractors. Additionally, a lesser-known approach utilizing Nambu mechanics\cite{PhysRevD.7.2405, Azc_rraga_1997, Takhtajan_1994, pandit1998generalized} - a generalization of Hamiltonian mechanics - offers an unique insights into the topological properties of these systems. This innovative method was first applied by P. Nevir and R. Blender \cite{nevir1994hamiltonian} in their analysis of the Lorenz attractor. This work was further extended by other researchers namely, M. Axenides and E. Floratos \cite{Axenides_2010, axenides2015scaling}, Z. Roupas \cite{Roupas_2012}, and W. Mathis and R. Mathis \cite{Mathis2013OscillatorSB, mathis2012stochastic}. But their work remain confided to only Lorenz, Rossler and other simple systems. This technique not only clarifies the intricate geometric structures \cite{Müller_2014, sahoo1993algebraic, Blender_2015} of chaotic attractors but also provides a refined understanding of phase space flows, as articulated by Liouville's theorem.

This Nambu mechanics is an active research topic in mathematical physics, particularly in systems with multiple conserved quantities and higher algebraic structure\cite{guha2002applications,10.1093/ptep/ptab075}. An  example of application of Nambu mechanics is : the Euler equations for a rigid body, which can be reformulated in Nambu mechanics with two Hamiltonians (energy and angular momentum squared) such as: $\frac{dL}{dt}=\qty\big{L,H,C}_{NB}$ where C=$\frac{1}{2}L^{2}$.

Nambu mechanics is a generalization of classical Hamiltonian mechanics proposed by Yoichiro Nambu in 1973\cite{PhysRevD.7.2405}. It extends the traditional Poisson bracket formalism to include multiple conserved quantities, leading to a higher-dimensional phase space dynamics (detail in section \ref{secIII}). Several key features of Nambu mechanics are : (i) Nambu-Poisson Bracket:
instead of the usual Poisson bracket $\pb{A}{B}_{PB}$ in Hamiltonian mechanics, Nambu mechanics introduces a triple bracket (or, more generally, an n-ary bracket) of the form  $\qty\big{A,B,C}_{NB}=\grad{A} \cdot (\grad{B} \cross \grad{C})$ where A,B,C are functions on phase space, and $\grad$ denotes the gradient in 3D phase space (x,y,z), (ii) Multiple Hamiltonians:
instead of a single Hamiltonian $H$, Nambu mechanics uses two (or more) quantities $H_1$, $H_2$  to define the time evolution, (iii) Volume-Preserving Flow: guiding principle of Nambu mechanics is the Liouville theorem. 	The dynamics in Nambu mechanics preserves the phase space volume (a generalization of Liouville’s theorem), analogous to how Hamiltonian mechanics preserves area in phase space, (iv) Applications: originally motivated by fluid dynamics \cite{guha2002applications}. Later extended to membrane physics (in String/M-theory)\cite{BAKER2001348} and constrained systems. Used in nonlinear dynamics and superintegrable systems\cite{doi:10.1063/1.1543227, ataguema2009generalization,10.1093/ptep/ptab075} where multiple conserved quantities exist. Also used for higher algebraic structures involved in Lie n-algebras and homotopy algebras\cite{2009}.

In this article, we explicitly demonstrated how different wings are produced by the intersections of Nambu hamiltonians surfaces. The key idea is to transform the governing equations into the form of gradient of these Nambu scalar functions. It is often not trivial to bring the original equations into the specific Hamiltonian forms. It requires some guess work and tricks. We implemented this approach of Nambu mechanics to a three-dimensional autonomous multi-scroll complex system, already studied numerically by L. Wang \cite{Wang2009} in this article. The reason for the choice of the known system is that it shows several number of scrolls depending on the choice of the system parameters. Therefore, this system serves as a compelling example of how the diverse geometries of Nambu surfaces can reveal the origins of multiple wings, all within a unified Nambu framework.

The mathematical model for multi-scroll system is stated in section \ref{secII}. An introductory idea about Nambu mechanics is mentioned in section \ref{secIII}. The importance of Nambu mechanics (in section \ref{secIVA}) and the geometrical findings from Nambu mechanics (in section \ref{secIVB}) are also discussed. Separate analyses are carried out for non-dissipative part and dissipative part in sections \ref{secIVB1} and \ref{secIVB2} respectively. In section \ref{secIVC}, we have illustrated the formation of n-scroll chaotic attractors for several set of parameters with the help of Nambu formalism. At last, we have concluded all the results of this article in section \ref{secV}. 


\section{\label{secII}A Multi-Scroll chaotic system }

The dynamical equations of a 3D quadratic autonomous system with three non-linear terms are given by 
\begin{equation}
\label{eq1}
\begin{cases}
\dot{x}=a(x-y)-yz \\
\dot{y}=-by+xz \\
\dot{z}=-cz+dx+xy 
\end{cases}\,.
\end{equation}
where (a, b, c, d) and (x, y, z) are system's control parameters and state variables respectively. These dynamical equations represent a well known multi-scroll system, which was introduced by L. Wang \cite{Wang2009}. 
We know that the chaotic systems can be broadly classified into conservative and dissipative systems based on their phase-space volume evolution and that can be calculated by the divergence of the system. Therefore, the divergence of this multi-scroll system is found to be

\begin{equation}
	\div {\vec v} =\frac{\partial \dot{x}}{\partial x}+ \frac{\partial \dot{y}}{\partial y}+\frac{\partial \dot{z}}{\partial z} =a-b-c<0
\end{equation}

\begin{figure}[htbp!]
{\includegraphics[width=0.99\linewidth]	{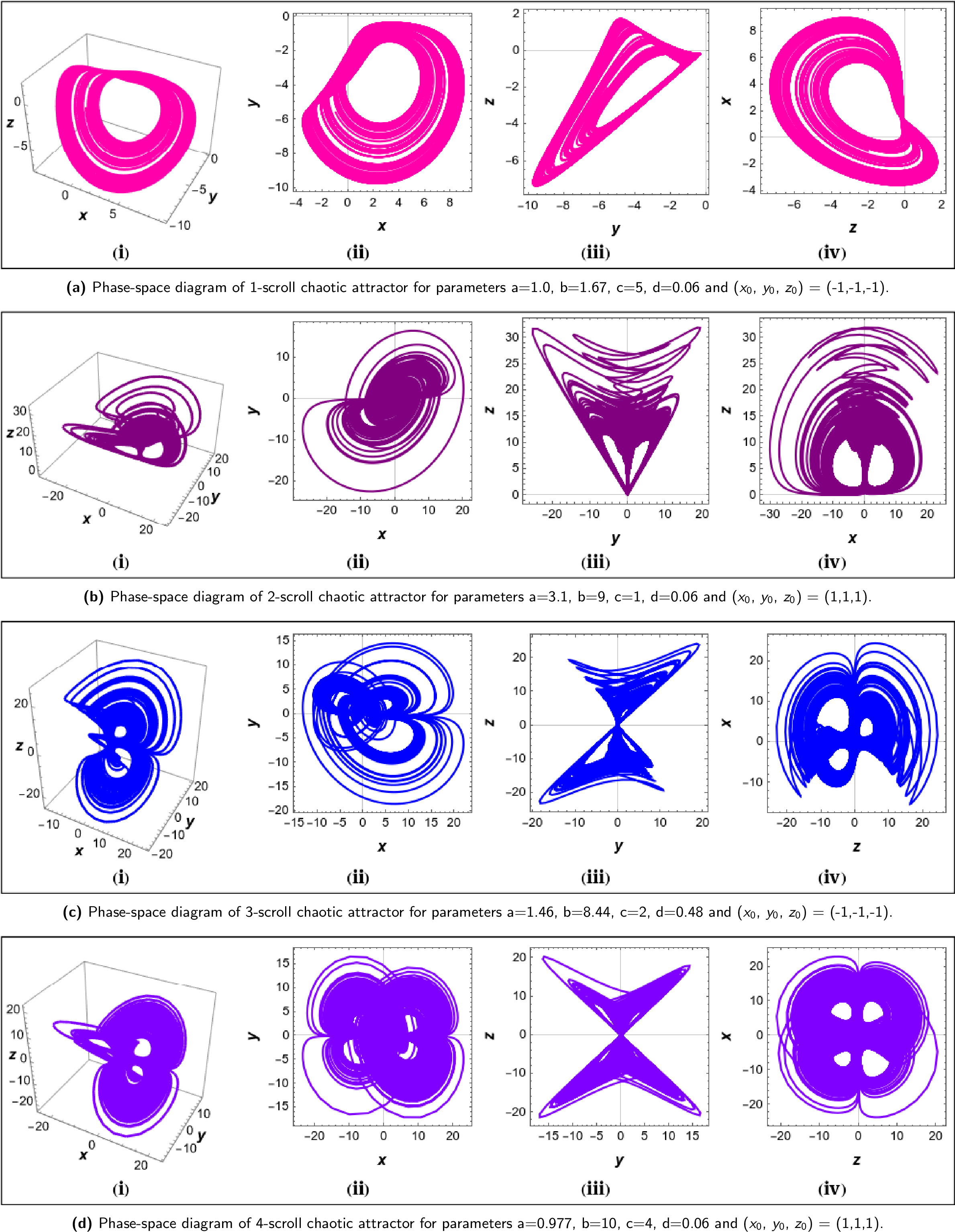}}

\caption{Phase-space trajectories of different multi-scroll attractors emerging from Eq.(\ref{eq1}) for various set of parameter values and initial conditions. In each sub-figure, (a-d)(i) represents 3D phase-space trajectories and (a-d)(ii), (a-d)(iii), (a-d)(iv) represents 2D phase-space trajectories in x-y plane, y-z plane, z-x plane respectively. }
\label{fig1}
\end{figure}

For dissipative systems  ($\div {\vec v} < 0$), meaning the volume of a region in phase space shrinks over time, leading to attractors (fixed points, limit cycles, or strange attractors). Hence, the dissipativity nature of the system(\ref{eq1}) confirms the existence of attractors. In the system under consideration, the $\div {\vec v}$ is always found to be negative for all four set of chosen parameter values (refer FIG. \ref{fig1}). 

 The system(\ref{eq1}) has five equilibrium points. These equilibrium points can be obtained by solving the equations ($\dot{x}=0$, $\dot{y}=0$, $\dot{z}=0$) and all five equilibria are given by

\begin{equation}
	\label{equili1}
	\begin{cases}
		S1= (0,0,0) \\
		S2= (\frac{-bd+\sqrt{(bd)^2+bc(-a+\sqrt{a^2+4ab})^2}}{-a+\sqrt{a^2+4ab}}, \frac{(-bd+\sqrt{(bd)^2+bc(-a+\sqrt{a^2+4ab})^2})}{2b}, \frac{-a+\sqrt{a^2+4ab}}{2})\\
		S3= (\frac{-bd-\sqrt{(bd)^2+bc(-a+\sqrt{a^2+4ab})^2}}{-a+\sqrt{a^2+4ab}},\frac{(-bd-\sqrt{(bd)^2+bc(-a+\sqrt{a^2+4ab})^2})}{2b},\frac{-a+\sqrt{a^2+4ab}}{2})\\
		S4= (\frac{-bd+\sqrt{(bd)^2+bc(-a-\sqrt{a^2+4ab})^2}}{-a-\sqrt{a^2+4ab}}, \frac{(-bd+\sqrt{(bd)^2+bc(-a-\sqrt{a^2+4ab})^2})}{2b}, \frac{-a-\sqrt{a^2+4ab}}{2})\\
		S5= (\frac{-bd-\sqrt{(bd)^2+bc(-a-\sqrt{a^2+4ab})^2}}{-a-\sqrt{a^2+4ab}}, \frac{(-bd-\sqrt{(bd)^2+bc(-a-\sqrt{a^2+4ab})^2})}{2b}, \frac{-a-\sqrt{a^2+4ab}}{2})
	\end{cases}\,.
\end{equation}
The Jacobian matrix of the system(\ref{eq1}) is found by linearizing the system around the equlibrium points and it is given by
\begin{equation}{\label{jac}}
	Jacobian, J=\mqty(a & (-a-z) & -y \\ z & -b & x \\ (d+y) & x & -c ) 
\end{equation}

We carried out numerical solution of the equation(\ref{eq1}). The phase space plot is shown in the FIG. \ref{fig1}. The FIG. \ref{fig1} indicates that the system produces various multi-scroll structures for different set of parameter values. Based on the literature and bifurcation diagram, we have considered four set of parameters such as $(a=1.0, b=1.67, c=5, d=0.06)$, $(a=3.1, b=9, c=1, d=0.06)$, $(a=1.46, b=8.44, c=2, d=0.48)$ and $(a=0.977, b=10, c=4, d=0.06)$ which give rise to 1,2,3,4-scroll chaotic attractors respectively. Apart from these values, many more set of parameters can be organized to produce more multi-scroll structures. The phase-space trajectories of different multi-scroll attractors emerging from equation(\ref{eq1}) is shown in FIG. \ref{fig1}. Depending on the correlation between parameters and the presence of coupled non-linear terms, the number of scrolls increases in this dynamical system. This increment in number of scrolls involved in the system, reflects the rising geometrical and dynamical complexity of the attractor. 

\begin{figure}[htbp!]
	{\includegraphics[width=0.99\linewidth]	{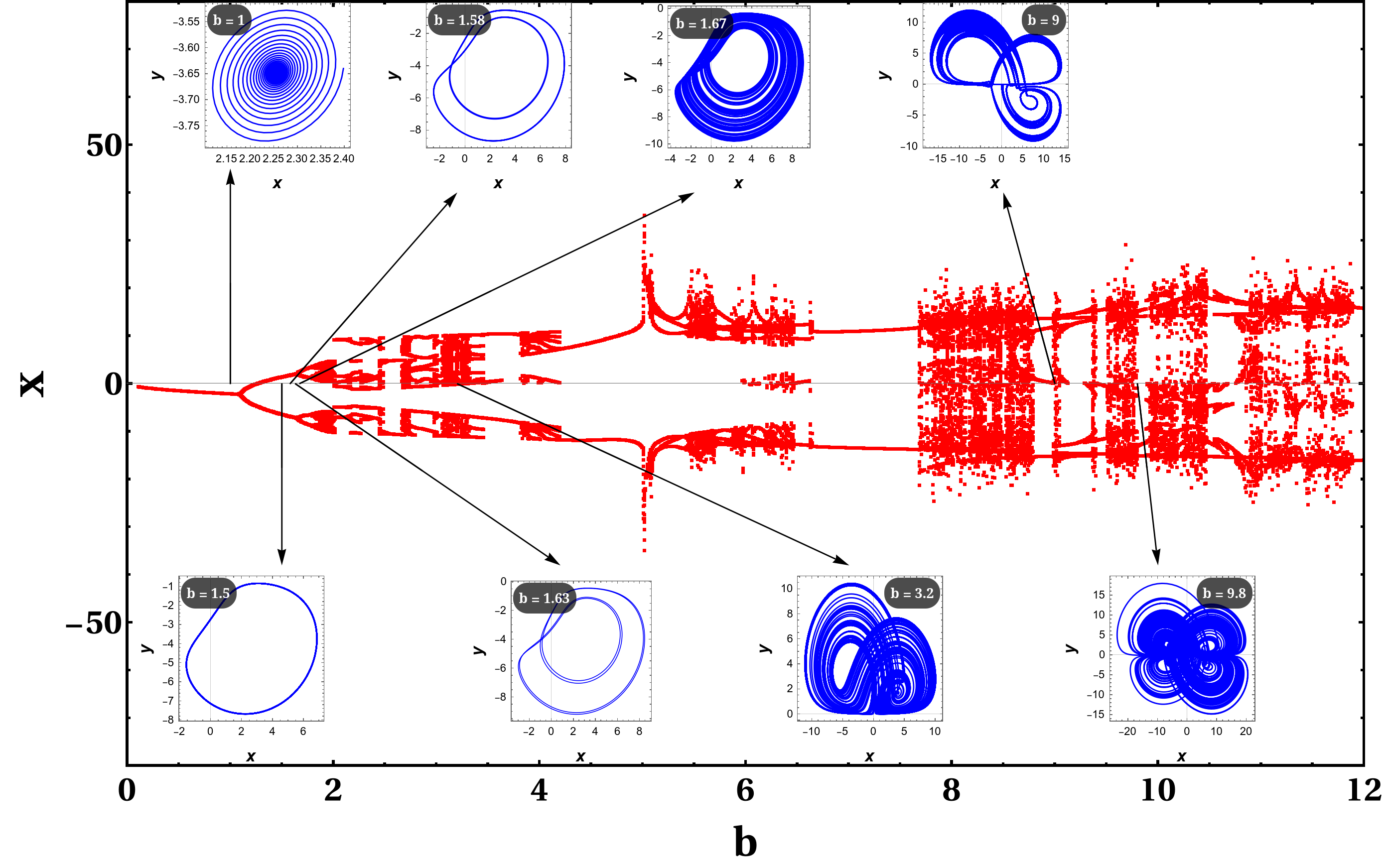}}
	
	\caption{It represents the bifurcation diagram, indicating the system's dynamics along x-axis for control parameter $“b”$ varying within the range $0 < b < 12$ by keeping $a=1, c=5, d=0.06$ fixed. And the inset figures illustrates the existence of multi-scroll attractors in the system(\ref{eq1}) by varying the control parameter $``b"$ such as a point attractor for $b=1$,  1-scroll limit cycle for $b=1.5$,  period-2 orbit for $b=1.58$,  period-4 orbit for $b=1.63$,  1-scroll chaotic attractor for $b=1.67$,  2-scroll chaotic attractor for $b=3.2$,  3-scroll chaotic attractor for $b=9$,  4-scroll chaotic attractor for $b=9.8$.}
		\label{figms}
\end{figure}

In order to confirm the evolution process of the dynamical system(\ref{eq1}), we have implemented bifurcation technique by varying parameter $``b"$ (as shown in FIG. \ref{figms}) to see the transitions between mono-stable and multi-stable regimes. The phase-space trajectories in $x-y$ plane, shown inside the FIG. \ref{figms}, depict routes to chaos in the system by varying the control parameter $``b"$ while keeping all other parameters at some fixed values such as $a=1$, $c=5$, $d=0.06$. Of course, similar phase-space plots illustrating the multi-stability of the system(\ref{eq1}) can be constructed if $a$ or $c$ or $d$ is chosen as the varying parameters. Conversely, we can state that the dynamical system's structural diversity is reflected in the multi-scroll attractor. Therefore, the existence of multi-stable chaotic attractors indicates that this multi-scroll dynamical system has extremely rich dynamics.
Using equation(\ref{jac}), we calculated the eigen values of the Jacobian matrix for all five equilibria of 1,2,3,4-scroll systems as shown in Table \ref{table1}.

\begin{figure}[htbp!]
	{\includegraphics[width=0.6\linewidth]	{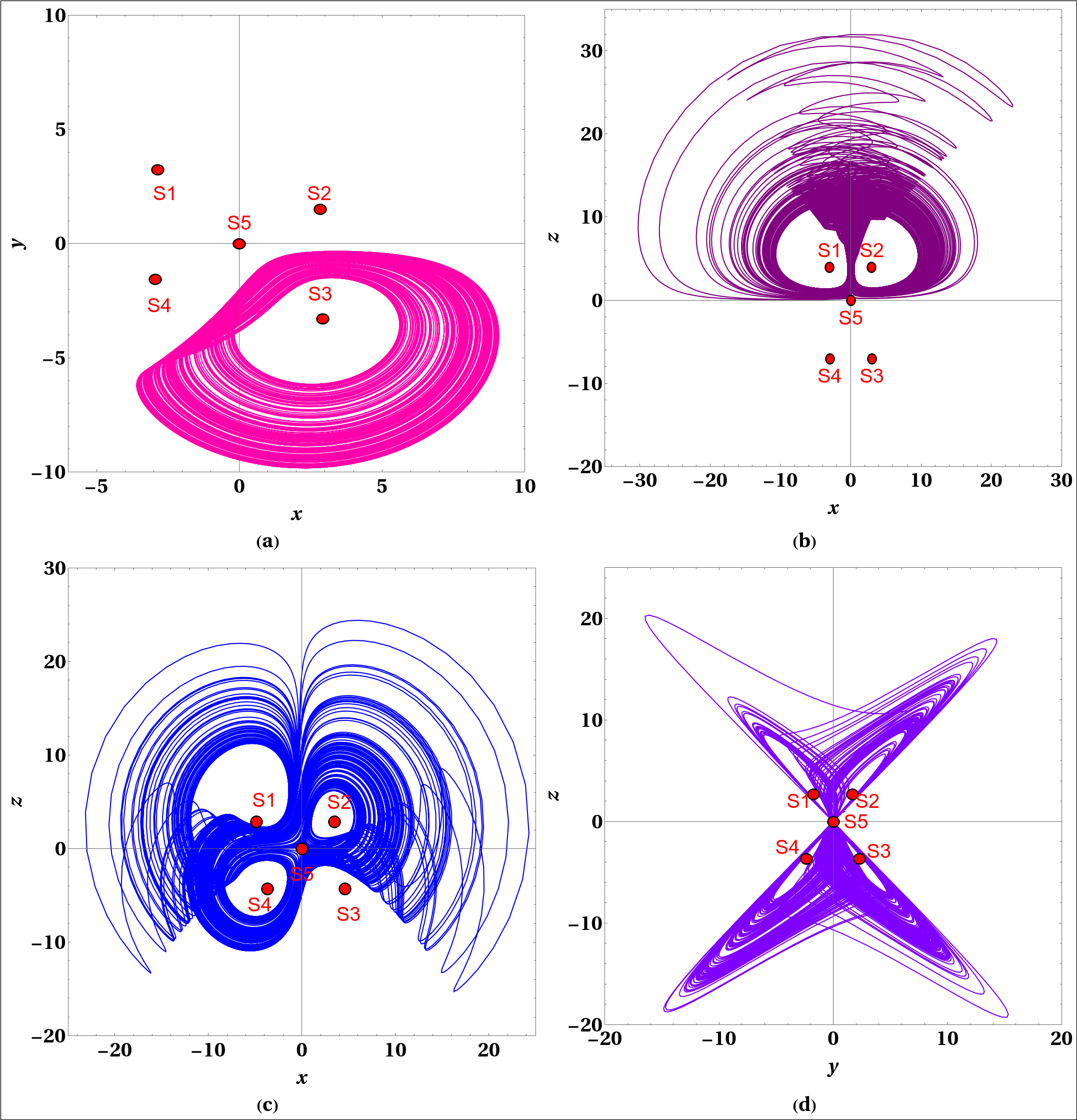}}
	
	\caption{It illustrates that the multi-scroll attractors have multiple equilibrium points and each "scroll" corresponds to a region around an  equilibrium point. Here, Red colour points indicate the equilibrium points of corresponding n-scroll attractor.}
	\label{eqpt}
\end{figure}

\begin{table}[htbp!]
	\centering
	\footnotesize
	\caption{\label{table1} The parameter values for 1, 2, 3, and 4-scroll system, their five equilibria, and their corresponding all three eigen values of the Jacobian matrix at each equilibrium points.}
	\begin{tabular}{l c l c l}
		\toprule[0.07cm]
		
		\bf{Systems} &   \bf{Parameters: $(a,b,c,d)$}& \bf{equilibrium points: $(x,y,z)$} & \bf{Eigen values of Jacobian: $\lambda1, \lambda2, \lambda3$} \\    
		
		\midrule

		\multirow{1}{*}{1-scroll}
		& $(1.0, 1.67, 5, 0.06)$& 
		
		$\begin{cases}
			(0, 0, 0) \\
			(-2.94676,-1.56274,0.885641) \\
			(-2.86319, 3.2329, -1.88564)\\
			(2.83362, 1.50274, 0.885641)\\
			(2.91633, -3.2929, -1.88564)
		\end{cases}$& 
		$\begin{cases} 
			-5, -1.67, 1 \\
			-6.67827 , 0.504136 \pm 1.86932 i \\
			-6.10855 , 0.219277 \pm 2.90321 i\\
			-6.5802 , 0.455102 \pm 1.85774 i\\
			-6.16657 , 0.248286 \pm 2.91395 i 
		\end{cases}$
		
		\\ \addlinespace

		\multirow{1}{*}{2-scroll}& 
		$(3.1, 9, 1, 0.06)$
		& 
		$\begin{cases}
			(0, 0, 0) \\
			(-3.06905, -1.3486, 3.95477)\\
			(-2.96197, 2.32178, -7.05477)\\
			(2.9325, 1.2886, 3.95477)\\
			(3.03852, -2.38178, -7.05477)
		\end{cases}$& 
		$\begin{cases} 
			-9, 3.1, -1 \\
			-8.3811 , 0.740551 \pm 3.17547 i\\
			-8.64501 , 0.872506 \pm 4.12081 i\\
			-8.26332 , 0.681658 \pm 3.13673 i\\
			-8.71814 , 0.909072 \pm 4.14991 i
		\end{cases}$
		
		\\ \addlinespace
		
		\multirow{1}{*}{3-scroll}&
		$(1.46, 8.44, 2, 0.48)$ 
		& 
		$\begin{cases}
			(0, 0, 0) \\
			(-4.87871, -1.65057, 2.85543)\\
			(-3.66587, 1.87438, -4.31543)\\
			(3.45993, 1.17057, 2.85543)\\
			(4.60464, -2.35438, -4.31543)
		\end{cases}$ & 
		$\begin{cases} 
			-8.44, -2, 1.46 \\
			-10.6037 , 0.811857 \pm 2.94085i\\
			-9.66038 , 0.34019 \pm 3.37511i\\
			-9.36002 , 0.190011 \pm 2.72799i\\
			-10.5126 , 0.766311 \pm 3.563 i
		\end{cases}$
		
		\\ \addlinespace  
		
		\multirow{1}{*}{4-scroll}& $( 0.977, 10, 4, 0.06 )$& 
		$\begin{cases}
			(0, 0, 0) \\
			(-6.43769, -1.72217, 2.67514)\\
			(-6.24295, 2.28001, -3.65214)\\
			(6.21341, 1.66217, 2.67514)\\
			(6.40723, -2.34001, -3.65214)
		\end{cases}$& 
		$\begin{cases} 
			-10, -4, 0.977 \\
			-13.914 , 0.445486 \pm 3.11596 i \\
			-13.8031, 0.390072 \pm 3.61512i\\
			-13.7059 , 0.341444 \pm 3.09694i \\
			-13.9576 , 0.467287 \pm 3.63327 i
		\end{cases}$
		
		\\ \addlinespace

		\bottomrule[0.07cm]
	\end{tabular}
\end{table}

From Table \ref{table1}, it is observed that out of five equilibrium points, four equilibrium points are index-2 saddle focus point and the other one equilibrium point is a index-1 saddle focus point(i.e, the central equilibrium point at $(x,y,z)=(0,0,0)$ ). An index -2 saddle focus is characterized by a single negative real eigenvalue and two complex conjugate eigenvalues, that possess positive real parts. These unique points serve as the dynamic centres for the  ``wings" or ``scrolls" that define the chaotic attractor\cite{8866023,HU2025826}. Their intricate behaviour not only highlights the complexity of the system but also emphasizes the beauty inherent in chaotic dynamics. And the index-1 point is crucial in defining the boundary between the distinct chaotic scrolls\cite{8866023}, effectively creating a separation that enhances our understanding of their complexities. So, it is confirmed that these four index-2 saddle focus equilibrium points are responsible for the generation of 4-wing/4-scroll chaotic attractor\cite{Jiang2020}. But the number of scroll/wing associated with the chaotic attractor changes depending on the parameter values. The same fact is verified by plotting the bifurcation diagram for the varying  parameter $``b"$ as shown in FIG \ref{figms} .

Therefore, except the central equilibrium point at origin, all other equilibrium points of the system is associated with a particular scroll/wing of the multi-scroll system, which is shown in FIG. \ref{eqpt} for several set of parameters (see Table \ref{table1} for corresponding parameter values) representing 1,2,3,4-scroll chaotic attractors.
\section{\label{secIII}Nambu Mechanics: The Triple Bracket Formulation }

The generalized Hamiltonian Mechanics is otherwise known as Nambu mechanics\citep{PhysRevD.7.2405}. In Hamiltonian mechanics, the equation of motions can be expressed in terms of Poisson bracket involving a pair of canonical variables or bivectors. In Nambu mechanics, the equations of motion are written in terms of the Poisson like Nambu brackets involving three or more canonical variables. Both Poisson bracket and Nambu bracket exhibits several properties such as antisymmetry, linearity, the Leibniz rule, the Jacobi identity etc.

In Hamiltonian mechanics, the equation of motion of a canonical variable in terms of Hamiltonian H using Poisson bracket is defined as
\begin{equation} 
\label{eq4} 
\dot{x_{i}}=\pb{x_{i}}{H}_{PB}=\epsilon_{ij}\partial_{j}{H}=\sum_{j} \epsilon_{ij}\frac{\partial{H}}{\partial{x_{j}}}
\end{equation}
where the Poisson bracket contains only a single Hamiltonian H and it is a constant of motion in conservative system.

In Nambu mechanics, the equations of motion of $n$ numbers of canonical variables involve ($n$-$1$) numbers of Hamiltonian functions $H_{1}$, $H_{2}$,...$H_{n-1}$ which can be represented as the Nambu bracket, similar to Poisson bracket. 
\begin{equation}
\label{eq10}
\dot{x_{i}} =\qty\big{{x_{i}},{H_{1}},{H_{2}}...{H_{n-1}}}_{NB}=\epsilon_{ijk...l}\partial_{j}{H_{1}}\partial_{k}{H_{2}}...\partial_{l}{H_{n-1}}=\sum_{jk...l} \epsilon_{ijk...l}\frac{\partial{H_{1}}}{\partial{x_{j}}}\frac{\partial{H_{2}}}{\partial{x_{k}}}...\frac{\partial{H_{n-1}}}{\partial{x_{l}}}
\end{equation}
For three canonical variables, the triple bracket representation of Nambu formulation is given by
\begin{equation}
\label{eq11}
\dot{x_{i}}=\qty\big{{x_{i}},{H_{1}},{H_{2}}}_{NB}=\epsilon_{qjk}\partial_{q}x_{i}\partial_{j}{H_{1}}\partial_{k}{H_{2}}=\epsilon_{ijk}\partial_{j}{H_{1}}\partial_{k}{H_{2}}=\sum_{jk} \epsilon_{ijk}\frac{\partial{H_{1}}}{\partial{x_{j}}}\frac{\partial{H_{2}}}{\partial{x_{k}}}=\grad{H_{1}}\cross\grad{H_{2}} 
\end{equation}

The equivalent Nambu-Poisson bracket for equation(\ref{eq11}) is given by

\begin{equation}\label{eq12}
\qty\big{x_{i},H_{1}}_{H_{2}}=\epsilon_{ijk}\partial_{j} H_{1}\partial_{k} H_{2}=\grad{H_{1}}\cross\grad{H_{2}}    
\end{equation}

In this Nambu-Poisson notation, the Hamiltonian written outside the Nambu-Poisson bracket represents a two-dimensional Euclidean space whereas the Hamiltonian mentioned inside the bracket denotes the system's dynamics on that two-dimensional Euclidean space. The role of $H_{1}$ \& $H_{2}$ can be interchanged by changing the positions of  $H_{1}$ \& $H_{2}$ in notation $\qty\big{x_{i},H_{1}}_{H_{2}}$. Hence, without taking the dissipation into account, the dynamical equations for three canonical variables in terms of Nambu bracket (by referring equations (\ref{eq11} and \ref{eq12})) is given by

\begin{equation}\label{NDeq}
\dot{\vec{x}} = \grad{H_{1}}\cross\grad{H_{2}}
\end{equation}

Here, both $H_{1}$ \& $H_{2}$ are constant of motions, because Nambu mechanics follows Liouville theorem as the guiding principle. It means this Nambu analysis can only be applied to those systems where phase space volume is preserved. In otherwords, we can say that the velocity field should be  divergenceless in Nambu mechanics. Therefore, equation(\ref{NDeq}) shows that there exists two Hamiltonians for a single 3D dynamical system. We can say that  the Nambu Hamiltonians form a doublet i.e. h=($H_{1},H_{2}$). One of the Nambu Hamiltonians represents a two-dimensional phase-space in $R^{3}$ and other Hamiltonian defines system's time evolution on this two-dimensional phase-space \cite{Takhtajan_1994, 2009}.

These Hamiltonian functions $H_1$ \& $H_2$ can be transformed to generate all other possible Nambu Hamiltonian functions by following a canonical transformation law , where the Jacobian(J) of the transformation matrix is equal to 1 and mathematically, such a transformation keeps the dynamics invariant. The condition is defined by

\begin{equation}
\label{condi}
\abs{\frac{\partial (H'_{1},H'_{2})}{\partial (H_{1},H_{2})}} =1
\end{equation}
Here, new transformed Hamiltonians $H'_1$ and $H'_2$ are the functions of $H_1$ and $H_2$. The transformation matrix can be any $2\cross2$ matrix in the $H_1$ and $H_2$ space, satisfying the condition (\ref{condi}). For any two transformed Hamiltonians, the equation (\ref{NDeq}) remains the same and the intersecting orbit of those $H'_1$ and $H'_2$ surfaces represents the same trajectories as that of $H_1$ and $H_2$ in the phase space.

The canonical transformation law for Nambu doublets is given by 
\begin{equation}
\label{trans}
\begin{split}
\grad{H'_{1}}\cross\grad{H'_{2}}&=\epsilon_{ijk}\partial_{j}{H'_{1}}\partial_{k}{H'_{2}}\\&
=\epsilon_{ijk}\qty\Bigg(\pdv{H'_{1}}{H_{1}}\partial_{j}{H_{1}}+\pdv{H'_{1}}{H_{2}}\partial_{j}{H_{2}})\qty\Bigg(\pdv{H'_{2}}{H_{1}}\partial_{k}{H_{1}}+\pdv{H'_{2}}{H_{2}}\partial_{k}{H_{2}})\\&
=\epsilon_{ijk}\qty\Bigg(\pdv{H'_{1}}{H_{1}}\pdv{H'_{2}}{H_{2}}-\pdv{H'_{1}}{H_{2}}\pdv{H'_{2}}{H_{1}})\partial_{j}{H_{1}}\partial_{k}{H_{2}}\\&
=\abs{\frac{\partial (H'_{1},H'_{2})}{\partial (H_{1},H_{2})}} (\grad{H_{1}}\cross\grad{H_{2}}) 
\end{split}
\end{equation}

This theory can only be applied to those systems that preserve the phase-space volume i.e. $\div{\vec{v}}=0$. Therefore the system where this condition is not satisfied like  3-D autonomous multi-scroll chaotic system, we need to decompose the whole system into two distinct parts namely, a conservative rotational non-dissipative part ($\vec v_{ND}$) and another non-conservative irrotational dissipative part ($\vec v_{D}$). Therefore, the system's flow vector is redefined as $\vec v_{flow}=\vec v_{ND}+\vec v_{D}$. The crucial conditions for this decomposition are : \quad $\div{\vec v_{ND}}=0$ (rotational)\quad and\quad $\curl\vec v_{D}=0$ (irrotational) . \\

We know that the Helmholtz-Hodge decomposition of a velocity field can be written as :
\begin{equation}
 \vec{v}=\curl\vec{A}+\vec\grad{D} \quad\Rightarrow \vec{v}=\vec v_{ND}+\vec v_{D}
\end{equation}

Hence,
\begin{equation}
\label{eq3}
\begin{cases}
\vec v_{ND}=\curl{\vec{A}}\quad\Rightarrow\div{\vec v_{ND}}=\div{(\curl{\vec{A}})}=0 \\
\vec v_{D}=\vec\grad{D}\quad\Rightarrow \vec\curl\vec v_{D}=\curl\grad{(D)}=0 
\end{cases}\,.
\end{equation}\\

where  $\vec{A}$ and D represents the vector field and  scalar field respectively.
\section{\label{secIV}Nambu Mechanics Approach to Multi-Scroll Chaotic System}

The application of Nambu mechanics to chaotic attractors is an emerging interdisciplinary topic that bridges generalized Hamiltonian dynamics and non-linear chaos theory. While Nambu mechanics is not yet a standard tool in chaos studies, its unique structure - particularly its multi-Hamiltonian and volume-preserving properties - offers intriguing possibilities for analysing and controlling chaotic systems like multi-scroll attractors.

\subsection{\label{secIVA}Advantage of Nambu mechanics framework for chaotic system}

We know, chaotic systems often possess multiple conserved quantities (e.g., energy, angular momentum), which align with Nambu’s multi-Hamiltonian formalism. The Nambu-Poisson bracket generalizes phase space dynamics, potentially capturing chaotic flows more naturally than classical Hamiltonian mechanics. Volume preservation in Nambu mechanics (generalizing Liouville’s theorem) matches the fractal phase-space structure of strange attractors.\\

The potential applications of Nambu mechanics to multi-Scroll attractors are :
\begin{itemize}
\item {Reformulating chaotic flows via Nambu brackets: consider a 3D chaotic system with two conserved quantities $H_1$, $H_2$.
The Nambu dynamics would be $\frac{dx}{dt}$= $\qty\big{x,H_1,H_2}_{NB}$ where the triple bracket defines the chaotic flow.}
\item {Controlling scroll generation: multi-scroll attractors rely on switching between equilibria. Nambu mechanics could provide a geometric control method by modulating $H_1$, $H_2$ .}

\item {Quantifying chaos with Nambu structures: classical Lyapunov exponents measure divergence in Hamiltonian systems. Nambu mechanics could generalize this via higher-order phase-space volumes.}

\end{itemize}

\subsection{\label{secIVB}Analysis of multi-scroll chaotic system}
In this article, we have implemented Nambu mechanics to study the multi-scroll system(\ref{eq1}). The main focus is to establish the connection between formation of specific number of wings and geometry of Nambu surfaces. Let us decompose the chaotic system into two distinct parts i.e. one is conservative non-dissipative($\vec v_{ND}$) part and another is a non-conservative dissipative ($\vec v_{D}$) part in such a way that

\begin{equation}
\label{eq16}
\vec{v}=\vec v_{ND}+\vec v_{D}\quad and \quad\div{\vec v_{ND}}=0\quad and\quad\curl\vec v_{D}=0
\end{equation}

Again,
\begin{equation}
\label{eq17}
\vec{v}=\vec v_{ND}+\vec v_{D}\quad \Rightarrow\:\; \dot\vec{r} = \vec\grad{H_{1}}\cross \vec\grad{H_{2}}+\vec \grad{D}
\end{equation}
such that 
\begin{equation}
\label{eq18}
\vec v_{ND}=\vec\grad{H_{1}}\cross \vec\grad{H_{2}}
\end{equation}
\begin{equation}
\label{eq19}
\vec v_{D}=\vec \grad{D}
\end{equation}
where, $H_{1}$ \& $H_{2}$ are two Nambu Hamiltonian functions and D is the dissipation function.\\

We identify the non-dissipative part as

\begin{equation}\label{nd}
\vec v_{ND}=(-ay-yz,  xz,  dx+xy)
\end{equation}

and dissipative part as

\begin{equation}\label{d}
\vec v_{D}=(ax, -by, -cz)
\end{equation}
The Nambu mechanics explains that the cross-product of the gradients of Nambu-Hamiltonian surfaces represents the intersection of these surfaces. As we know, the gradients $\vec\grad{H_{1}}$ and $\vec\grad{H_{2}}$ are normal to the Hamiltonian surfaces $H_{1}$ and $H_{2}$ and the cross-product $\vec\grad{H_{1}}\cross \vec\grad{H_{2}}$ gives a vector perpendicular to both the normals $\vec\grad{H_{1}}$ and $\vec\grad{H_{2}}$. Hence, this cross-product vector(which is nothing but a velocity vector as shown in \ref{eq18}) should be tangent to both the surfaces $H_{1}$ and $H_{2}$ simultaneously, and hence tangent to their curve of intersection. In other words, we can say that this vector represents the dynamics which naturally follow this intersecting region of both the surfaces. Therefore,the  trajectories must lie on both the surfaces simultaneously. \\
		 
These surfaces $H_{1}$ and $H_{2}$ are conserved quantities i.e. $\frac{d H_{1}}{dt}= 0$ and $\frac{d H_{2}}{dt}= 0$ .This means the system is constrained to move on the intersection of the surfaces $H_{1}=k1$ and $H_{2}=k2$. These Hamiltonian surfaces are 2D surfaces in 3D-space and their intersection is typically a 1D-curve. This makes Nambu mechanics a geometrically natural formulation where the conservation laws and allowed trajectories are unified in a single mathematical structure. And the cross-product is not just a convenient choice - it is the only vector that points along the intersection of constraint surfaces. 

Let us perform a detail analysis about the contribution of non-dissipative part and dissipative part for the construction of multi-scroll chaotic attractors separately in forthcoming sub-sections.
\subsubsection{\label{secIVB1}Non-dissipative part analysis}
Nambu Hamiltonians ($H_{1}$, $H_{2}$) for this system are found to be 

\begin{equation}
\label{eq20}
\begin{cases}
H_{1}=z^{2}-(y+d)^{2} \\ 
H_{2}=\frac{1}{4}(x^{2}+y^{2})+\frac{1}{2}a z-\frac{1}{2}ad \log \abs{z+y+d}


\end{cases}\,.
\end{equation}
These two Hamiltonian functions $H_{1}$ \& $H_{2}$ are derived from the conservative non-dissipative part such that they satisfy the equation(\ref{eq18}). Hence, both $H_{1}= H_{1}(\vec r(0))$ \& $H_{2}= H_{2}(\vec r(0))$ represent conserved quadratic surfaces and the equation(\ref{eq20}) govern the evolution of the multi-scroll system(\ref{eq1}) over quadratic surfaces $H_1$ \& $H_2$ when the dissipation part is absent. Here, $H_1$ represents a hyperbolic surface where as $H_2$ represents a deformed parabolic surface. Since they are formed as a pair from the same equation, they are also known as Nambu doublet.

The intersection of two Nambu surfaces defined in equation(\ref{eq20}) represents the non-dissipative trajectory of the system, which is constructed using the following equation:
\begin{equation}\label{sur1}
	H_1 = k1 \quad \& \quad
	H_2 = k2
\end{equation}

where $H_{1}$ and $H_{2}$ are the two Nambu-Hamiltoian quadratic surfaces at the same initial condition.

Therefore, the intersecting orbit of $H_1 = k1$ and $H_2 = k2$ both at the same initial condition must give the non-dissipative phase-space trajectory of the corresponding n-scroll chaotic attractor (detail analysis is shown in section \ref{secIVC}). From the perspective of topology, every geometric shape has a corresponding energy. As a result, there is some energy associated with these Nambu-Hamiltonian surfaces. This indicates that the expansion of the hyperboloid surface's upper and lower halves and the deformed paraboloid surface both show the relevant energy connected to these geometrical structures. The initial conditions quantify those energy.\\


These two Nambu doublet(\ref{eq20}) can be transformed using the transformation law given by equation(\ref{trans}), where the Jacobian(J) of the transformation matrix is equal to 1. All the possible transformed Nambu surfaces obtained from transformation law are qualitatively same, but their axes of orientation may not be same. Only four types of Nambu surfaces are possible, namely hyperboloid, paraboloid, ellipsoid and cylindrical. The intersecting orbits of any two transformed Nambu surfaces are same, which is equivalent to the non-dissipative orbit of the chaotic system. This fact has already been proved by Z. Roupas\cite{Roupas_2012} while studying the Lorenz attractor. We have also verified the same results in our recent paper\cite{patra2025phasespacegeometryfourwings} while analysing the four-wing attractor.
\subsubsection{\label{secIVB2}Dissipative part analysis }
From equation(\ref{d}), we concluded that dissipative part of multi-scroll system involves linear terms only. Hence, the alternate and more concise mathematical representation of the dynamical equations(\ref{eq17}) is given by

\begin{equation}
\label{alter}
\begin{aligned}
\dot{x}=\left(\vec\grad{H_{1}}\cross\vec\grad{H_{2}}\right)_{1} -{\eta}_1 x\\
\dot{y}=\left(\vec\grad{H_{1}}\cross\vec\grad{H_{2}}\right)_{2} -{\eta}_2 y\\
\dot{z}=\left(\vec\grad{H_{1}}\cross\vec\grad{H_{2}}\right)_{3} -{\eta}_3 z\\
\end{aligned}
\end{equation}
where  $\eta_1 = -a$, $\eta_2 = b$ \& $\eta_3 = c$. This equation(\ref{alter}) defines the dynamics of whole multi-scroll system at different time instants due to linear dissipation. 
New variables ($u$, $v$, $w$, $\tau$) are defined in terms of the old variables ($x$, $y$, $z$, $t$) as :\\
\begin{equation}
u = e^{\eta_1 t}x,\quad v = e^{\eta_2 t}y,\quad  w = e^{\eta_3 t}z ,\quad \tau = \frac{1}{\eta_1 + \eta_2+ \eta_3} e^{\left(\eta_1+ \eta_2 + \eta_3\right)t}
\end{equation}

This dynamical system in terms of new variables again can be represented in an mathematical expression equivalent to the equation(\ref{eq18}), which is given by :
\begin{equation}
\label{equiv}
\frac{d}{d\tau}(u,v,w)= \vec\grad_{\vec q} {H_{1}}\cross\vec\grad_{\vec q}{H_{2}}
\end{equation}

where $\vec q \equiv u,v,w$ and explicitly time dependent Hamiltonians are $H_{i} = H_{i}\left(e^{-\eta_1 t}u, e^{-\eta_2 t}v, e^{-\eta_3t}w\right)$ , $i = 1$, $2$ . Since equation(\ref{equiv}) is equivalent to the equation(\ref{eq18}), the time derivative of (u, v, w) denotes phase space volume elements and according to the principles of Nambu mechanics, the phase-space consisting of ($u$, $v$, $w$) is divergenceless. The only difference between equation(\ref{equiv}) and equation(\ref{eq18}) is that the former equation, in terms of new variables, defines non-dissipative trajectories at different instants and as a whole, alone produces the full chaotic trajectories of the multi-scroll system as a collection of intersecting orbits of continuously deforming surfaces.\\

As we know, the part of the system that preserving phase-space volume (or flows)  are called "non-dissipative" part of the system and the non-conserving flows are called "dissipative" part of the system. And the dissipation function, D can be derived from equations (\ref{d}) and (\ref{eq19}), which is given by
\begin{equation}
	D=\frac{1}{2}(ax^{2}-by^{2}-cz^{2})
\end{equation}
 
In another way, we can say that when the geometrical power of the $H_{1}$ \& $H_{2}$ surfaces in this Nambu mechanics  is lost ( i.e. for D $\neq$ 0 ), then the first case what comes into our mind is that $H_{1}$ \& $H_{2}$ are not conserved anymore. Hence, $\frac{d H_{1}}{dt}\neq 0$ and $\frac{d H_{2}}{dt}\neq 0$. 
And  the time evolution equations are given by 
\begin{equation}
	\begin{cases}
	\dot{H_{1}}=\vec\grad{H_{1}}\vdot\vec v
	=\vec\grad{H_{1}}\vdot(\vec v_{ND}+\vec v_{D})
	=\vec\grad{H_{1}}\vdot\vec v_{ND}+ \vec\grad{H_{1}}\vdot \vec v_{D}
	=\vec\grad{H_{1}}\vdot\vec\grad{D}\\
	
	 \dot{H_{2}}=\vec\grad{H_{2}}\vdot\vec v
	=\vec\grad{H_{2}}\vdot(\vec v_{ND}+\vec v_{D})
	=\vec\grad{H_{2}}\vdot\vec v_{ND}+ \vec\grad{H_{2}}\vdot \vec v_{D}
	=\vec\grad{H_{2}}\vdot\vec\grad{D}\\
	
	 \dot{D}=\vec\grad{H_{2}}\vdot\vec v
	=\vec\grad{D}\vdot(\vec v_{ND}+\vec v_{D})=\vec\grad{D}\vdot\vec v_{ND}+ \vec\grad{D}\vdot \vec v_{D}=\qty\big{D,H_{1},H_{2}}+(\vec\grad{D})^{2}
	\end{cases}\,
\end{equation}

where,
\begin{equation}
 \vec\grad{H_{1}}\vdot\vec v_{ND}=0 ,\quad \vec\grad{H_{2}}\vdot\vec v_{ND}=0 ,\quad \vec\grad{D}\vdot\vec v_{ND}=\vec\grad{D}\vdot (\vec\grad{H_{1}}\cross \vec\grad{H_{2}})=\qty\big{D,H_{1},H_{2}}\end{equation}

Henceforth, when the dissipative part of the system comes into account, we found that $H_{1}$ and $ H_{2}$ are not conserved  for the full multi-scroll chaotic system and there is an irregular motion of the two surfaces against each other and their intersection varies with time. We also noticed that the system jumps from one periodic orbit to other periodic orbit of the non-dissipative sector at every moment. Therefore, the superposition of intersecting orbits of continuously deforming surfaces gives rise to the phase-space dynamics of the full chaotic multi-scroll system. And this is how the whole chaotic dynamics comes into picture due to Nambu mechanics.

Finally, our volume preserving system is given by the time derivative of $H_{1}$, $ H_{2}$ and $D$, which is given below:
\begin{equation}
	\begin{cases}
		\dot H_{1}= 2z \dot z - 2(y+d) \dot y\\
		\dot H_{2}=\frac{1}{2}(x \dot x + y \dot y)+ \frac{1}{2} a \dot z - \frac{1}{2} a d \frac{(\dot y + \dot z)}{(z+y+d)}\\
		\dot D= a x \dot x - b y \dot y - c z \dot z
	\end{cases}\,.
\end{equation}

Therefore, we can say that we cannot produce a dissipative chaotic attractor using the fundamental, volume-preserving principles of Nambu mechanics alone. In other words, we can state that it is impossible to get the ``attraction" part of the ``attractor" with pure Nambu mechanics. However, we can produce a system that has a chaotic attractor by using a ``Nambu-like" formulation where the equations of motion are split into a conservative part generated by a Nambu bracket and a dissipative part. This approach provides a beautiful geometric interpretation of chaos, where the combined effect of conservative energy mixing and dissipative energy loss gives rise to complex, chaotic dynamics on a strange attractor.
\subsection{\label{secIVC} Emergence of multi-scroll attractors from Nambu Doublet}

\begin{figure}[htbp!]
	\centering 
	\subfigure[\label{}]{\includegraphics[width=0.38\linewidth] {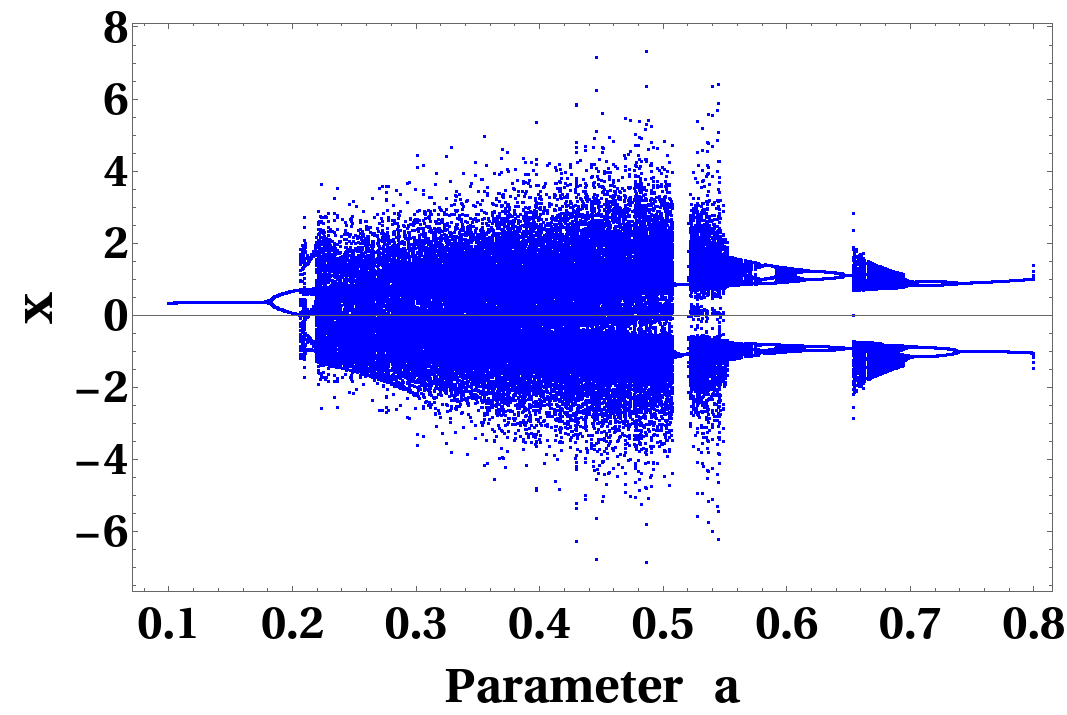}}
	\hfil
	\subfigure[\label{}]{\includegraphics[width=0.38\linewidth] {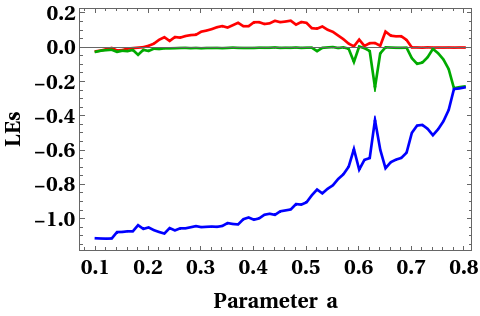}}
	\vfill
	\subfigure[\label{}]{\includegraphics[width=0.38\linewidth] {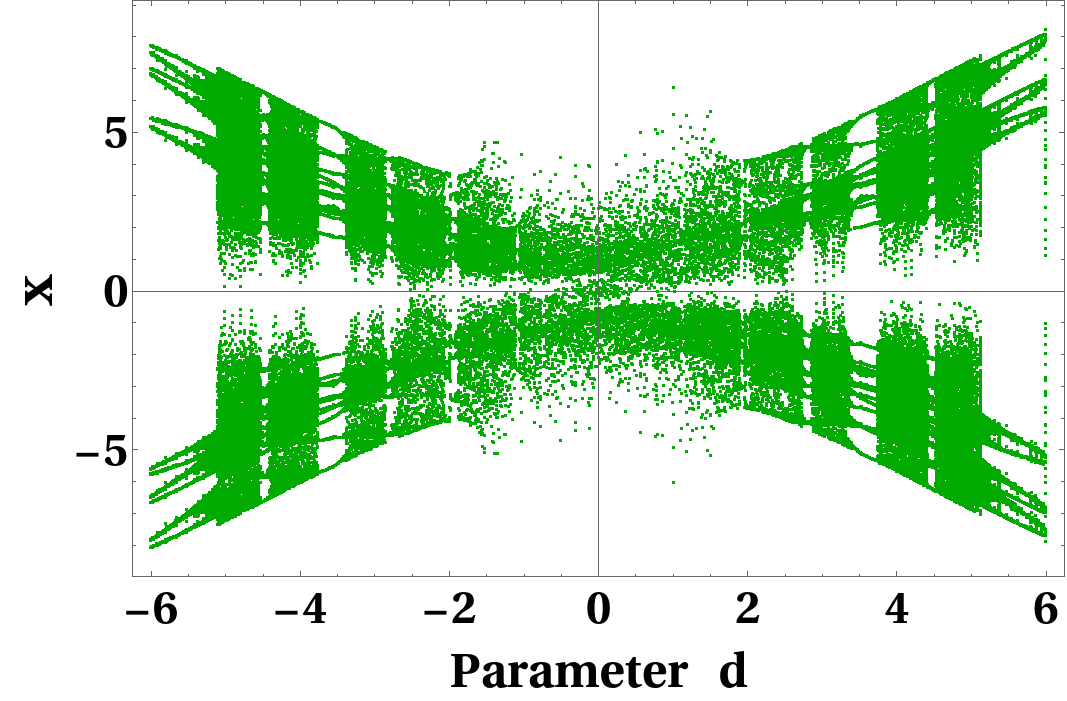}}
	\hfil
	\subfigure[\label{}]{\includegraphics[width=0.38\linewidth] {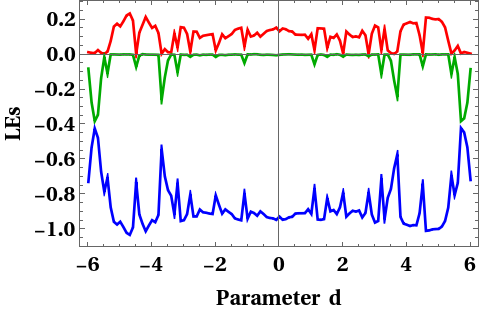}}
	
	\caption{Bifurcation diagrams in (a) and (c) indicate the existence of corresponding chaotic regimes for the individual variation of control parameter ``a'' and ``d'' respectively. The corresponding Lyapunov exponent spectrums are shown in (b) and (d).}
	\label{figbif}
\end{figure}

\begin{figure}[htbp!]
	{\includegraphics[width=0.4\linewidth]	{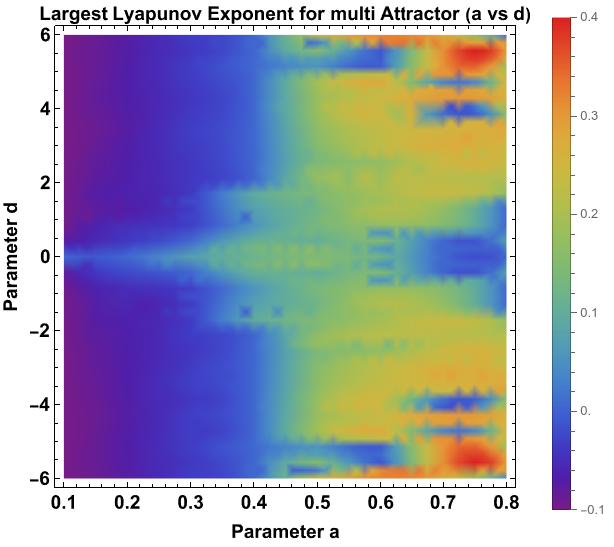}}
	
	\caption{Density plot of maximum Lyapunov exponents in a-d parameter space.}
	\label{figdensityplot}
\end{figure}

\begin{figure}[htbp!]
	\centering 
	\subfigure[\label{}]{\includegraphics[width=0.99\linewidth] {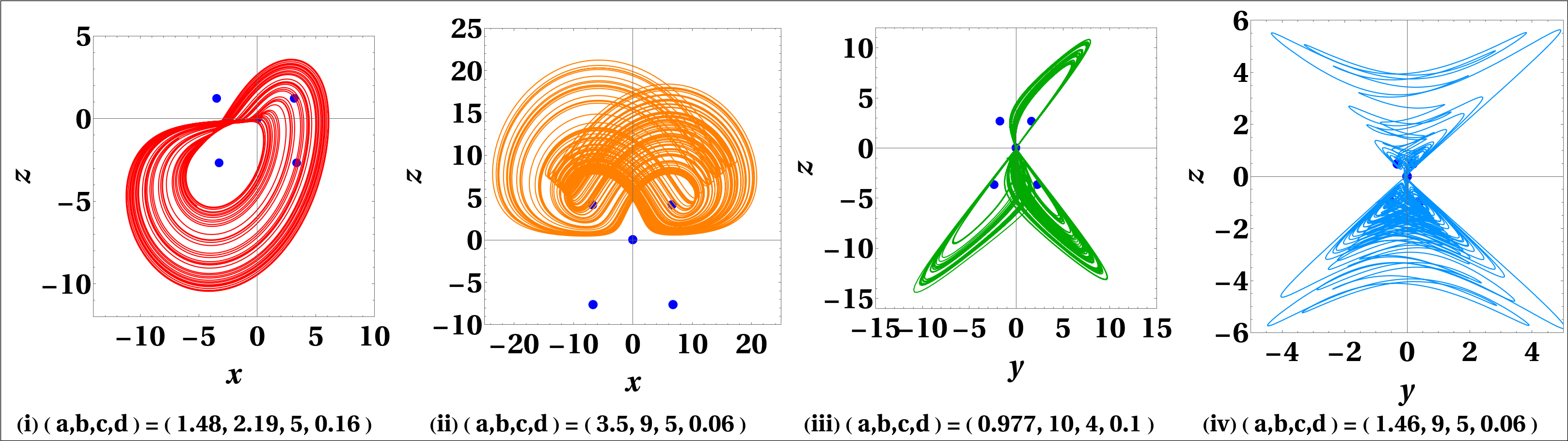}}
	\vfil
	\subfigure[\label{}]{\includegraphics[width=0.99\linewidth] {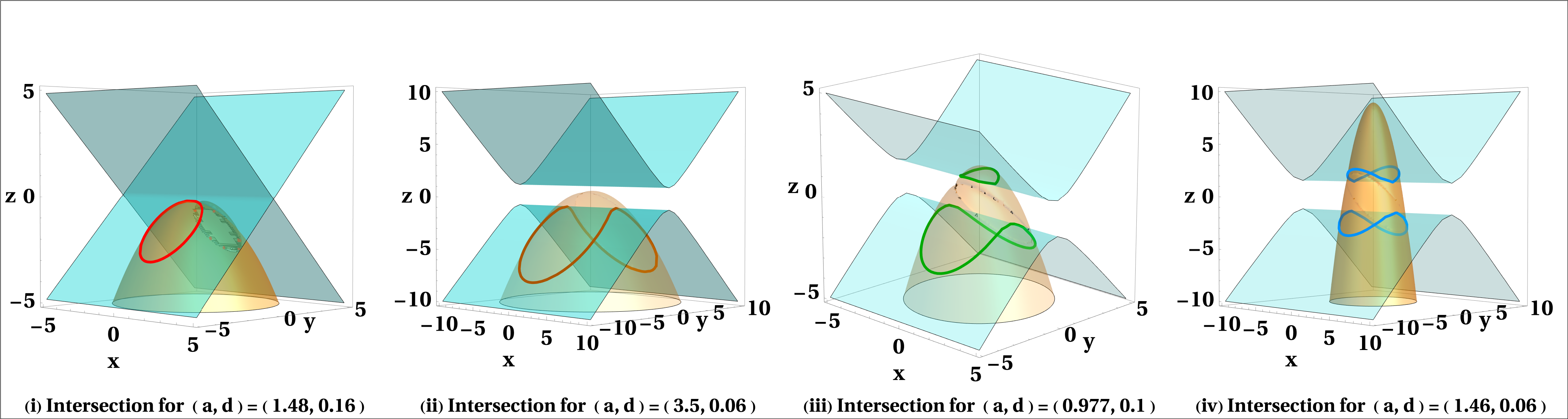}}
	\vfill
	\subfigure[\label{}]{\includegraphics[width=0.99\linewidth] {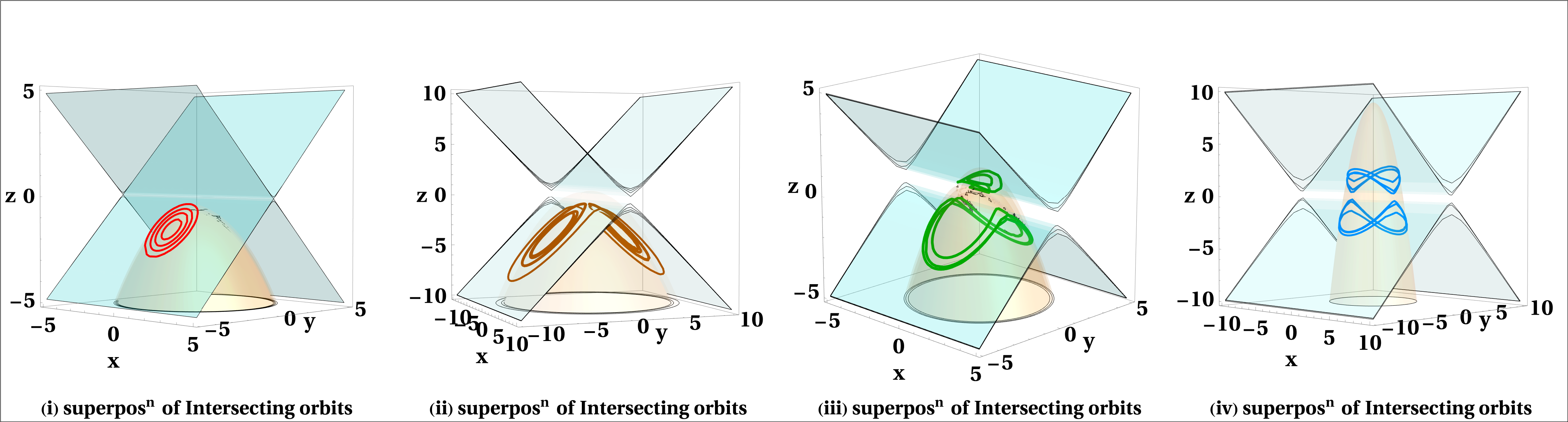}}

	\caption{Here, a(i-iv) represent 1,2,3,4-scroll chaotic attractors in phase space for four different combination of (a,b,c,d)- values and 1,2,3,4-scroll attractors are shown in red, orange,green, sky blue colour and blue colour dots represent their respective equilibrium points. The  intersecting orbit of two Hamiltonian surfaces for the corresponding (a,d)-parameters are illustrated in b(i-iv) and the superpositions of a number of intersecting orbits producing full 1,2,3,4-scroll attractor is shown in c(i-iv). }
	\label{IOs}
\end{figure}

A system can be integrable or chaotic depending on parameters. Hence, we need to choose corresponding suitable values of the parameters ($a$, $b$, $c$, \& $d$) to generate Nambu surfaces for different n-scrolls. For this purpose we plot bifurcation and Lyapunov exponent plots to obtain the suitable chaotic regimes.  The bifurcation diagram and Lyapunov exponents are shown in the FIG. \ref{figbif}. We consider parameters ``$a$" and ``$d$" both as control parameters in these plots because the Nambu doublet involve only these two parameters in their expressions(\ref{eq20}). Based on the available literature\cite{Wang2009}, we set the range ($0.1<a<0.8$) and ($-6<d<6$), while keeping other two variables fixed at $b=1$, $c=0.256$.  Setting $H_{1}=k1$ \& $H_{2}=k2$ with values of $k1$, \& $k2$ at certain initial condition, a non-dissipative orbit of n-scroll attractor is produced. In other words, we can say that for a pair of values of (k1, k2) at a particular initial condition, the corresponding intersecting orbit indicates the dynamics of the system at a particular instant corresponding to that initial time. Here, both the doublet are considered at the same initial condition.\\

 Notice that bifurcation diagrams and Lyapunov exponents (refer FIG. \ref{figbif}) can confirm the chaotic dynamics of the multi-scroll system in certain regimes, but they can not confirm the exact number of scrolls involved in the emerging chaotic attractors. Of  course, the concerned number of scrolls can be confirmed from phase-space plots only. All the numerical calculations are done using Mathematica 14.2. LE is calculated using Gram-Schmidt method.\\

 The density plot of LEs in $a$-$d$ parameter space indicates the possible chaotic regions for all combination of ($a$, $d$) parameters in a certain range. The colour bar in the density plot shows the magnitude of maximum LE involved with different coloured regions. As we can see from FIG. \ref{figdensityplot}, more than half of the regions of this density plot are involved with the dynamics having some positive LE. Thus the multi-scroll system possesses various n-scroll chaotic attractors within that certain range and accordingly the complexity (and magnitude of positive LE) of the system increases with the number of scrolls involved. The system is more chaotic within the range a$\in$[$0.6$ ,$0.8$]. There exist a very small window shown in violet colour existing in the left side of the density plot, for which the maximum LE is not positive, indicating the absence of chaos.

Out of all the possible combinations of ($a$, $d$), we consider those values of ($a$, $d$)-parameters which generate various intersecting orbits of $H_{1}=k1$ \& $H_{2}=k2$ at some instant leading to different number of scroll dynamics, producing different number of wings geometry in the phase space.

Let us generate different n-scroll attractors for different values of a and d parameters. Because, from equation (\ref{eq20}), we concluded that formation of n-scroll orbit is only possible when we change both parameter a and d simultaneously, irrespective of any value of parameter b and d. But, we should keep this in mind that as a whole, for any combination of $(a,b,c,d)$, the divergence of total vector field should be -ve.  It means as a whole those set of parameter values should produce a dissipative dynamical system.

\begin{figure}[htbp!]
	\centering 
	\subfigure[\label{}]{\includegraphics[width=0.49\linewidth] {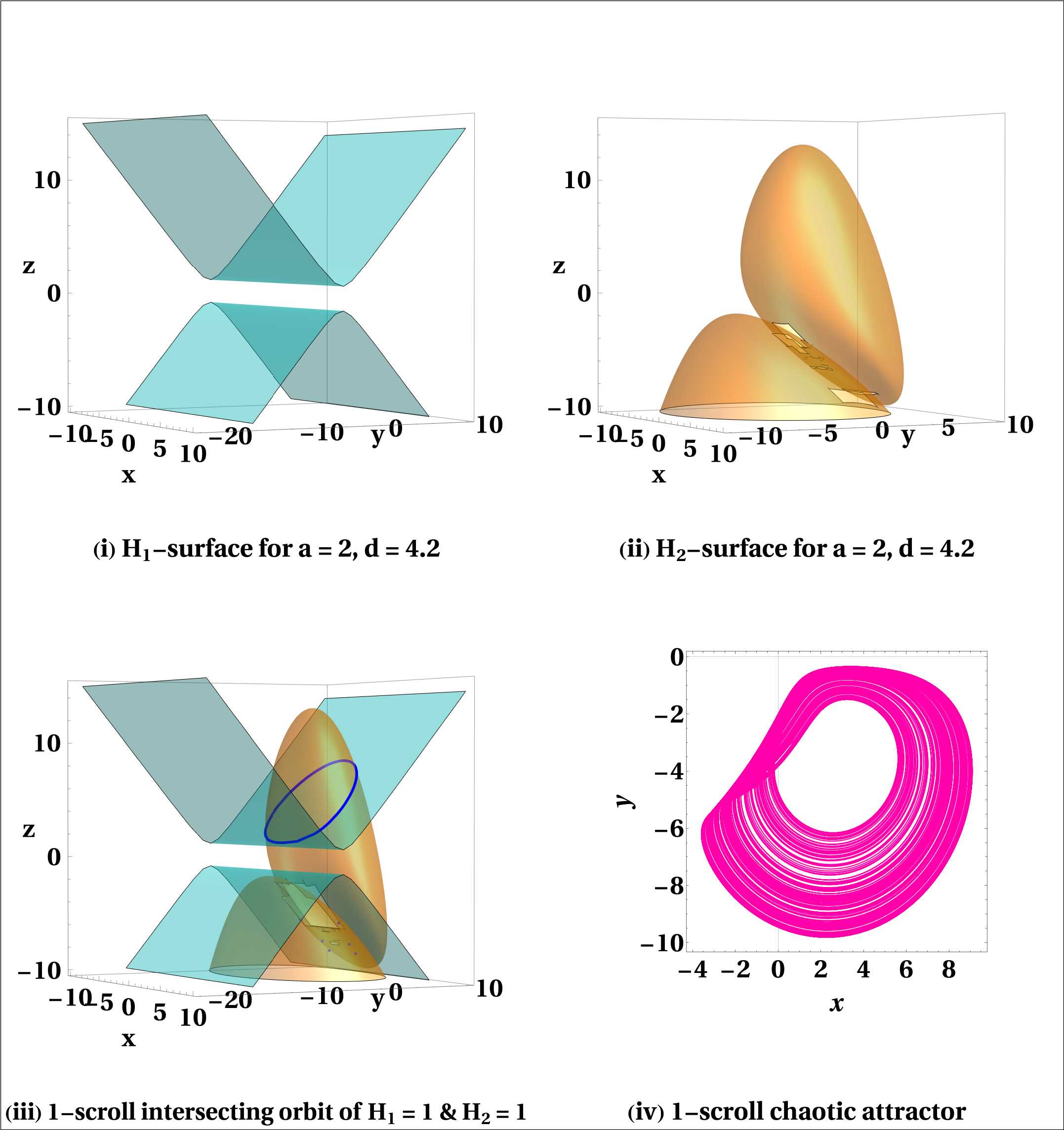}}
	\hfil
	\subfigure[\label{}]{\includegraphics[width=0.49\linewidth] {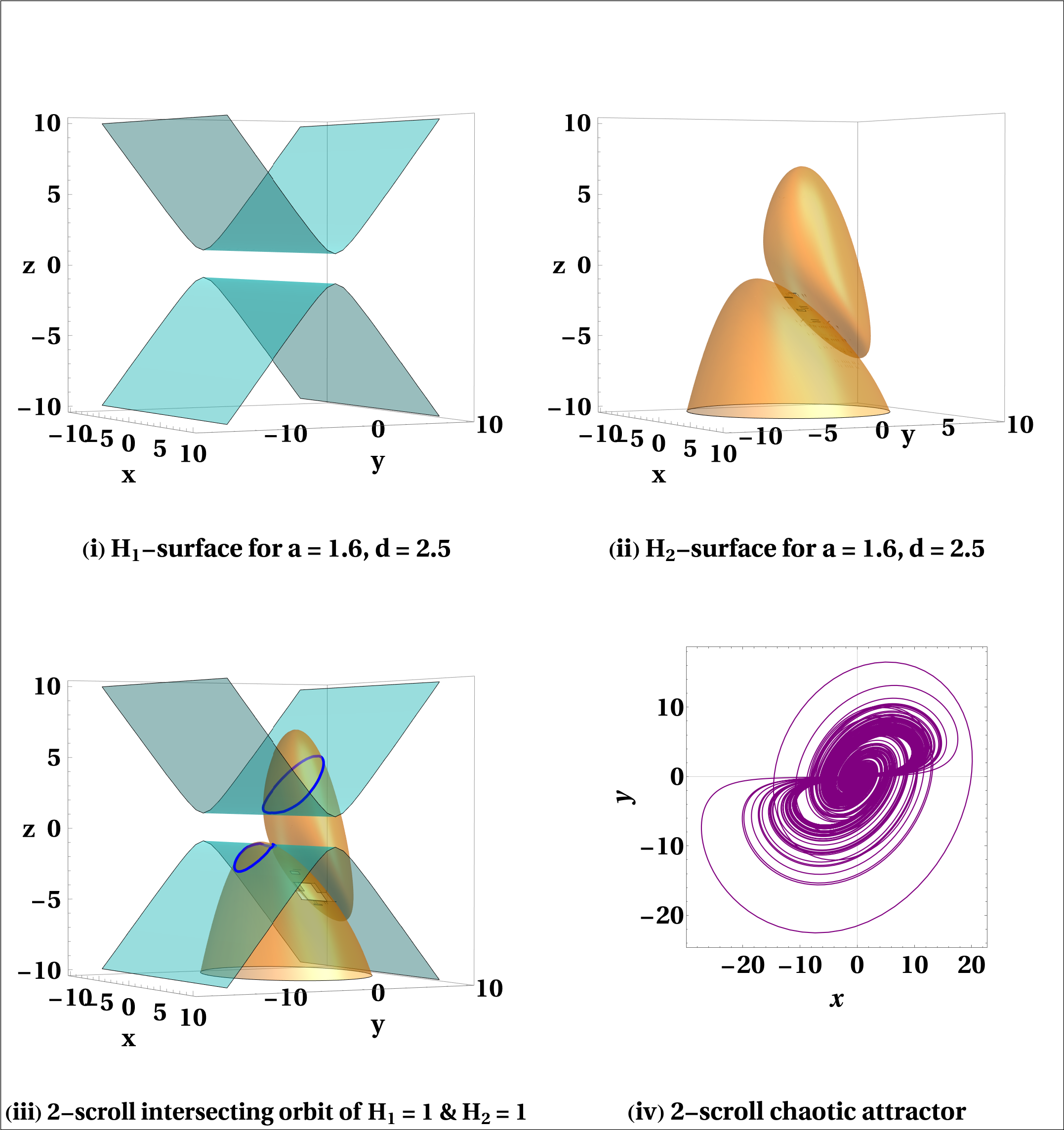}}
	\vfill
	\subfigure[\label{}]{\includegraphics[width=0.49\linewidth] {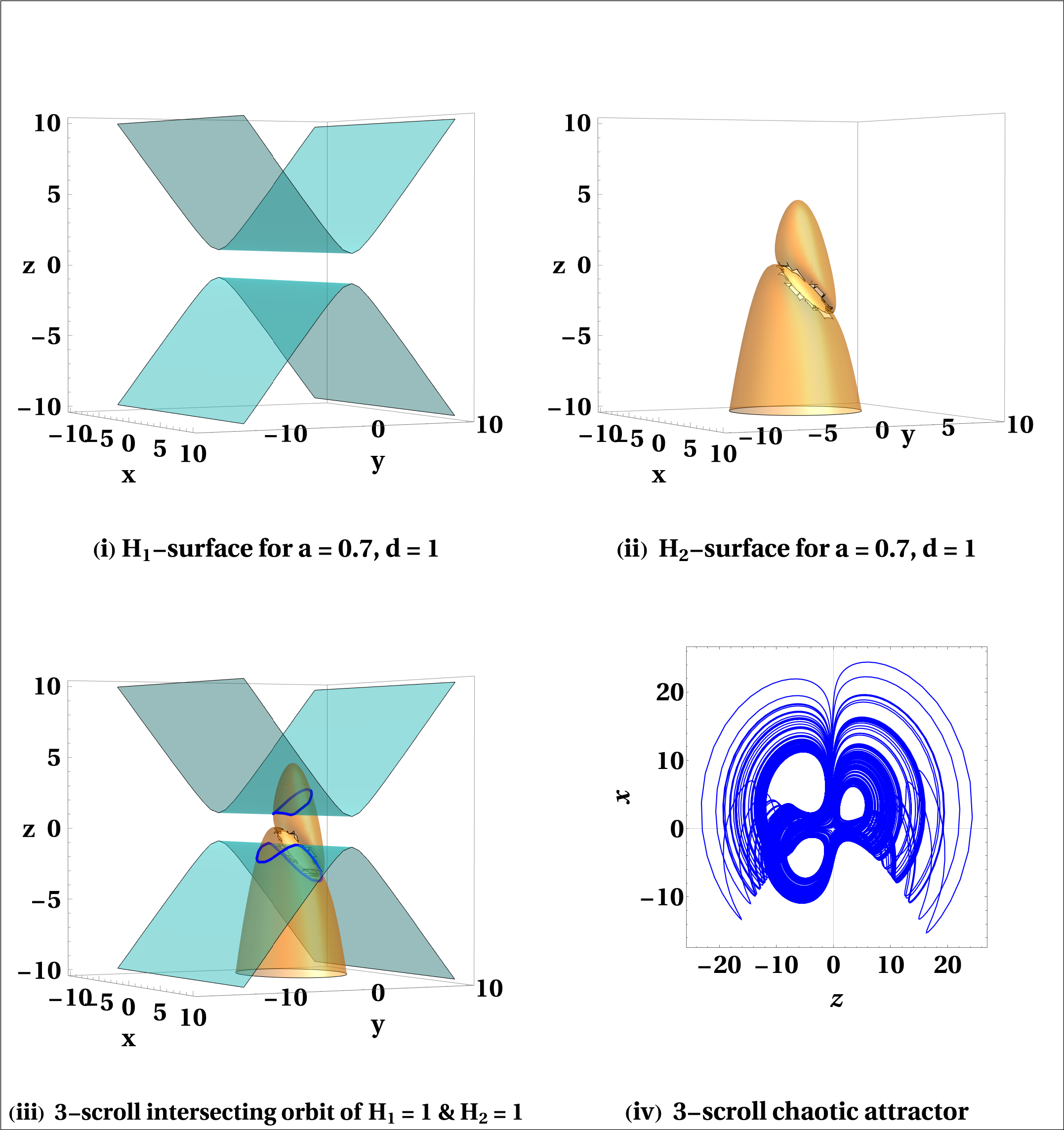}}
	\hfil
	\subfigure[\label{}]{\includegraphics[width=0.49\linewidth] {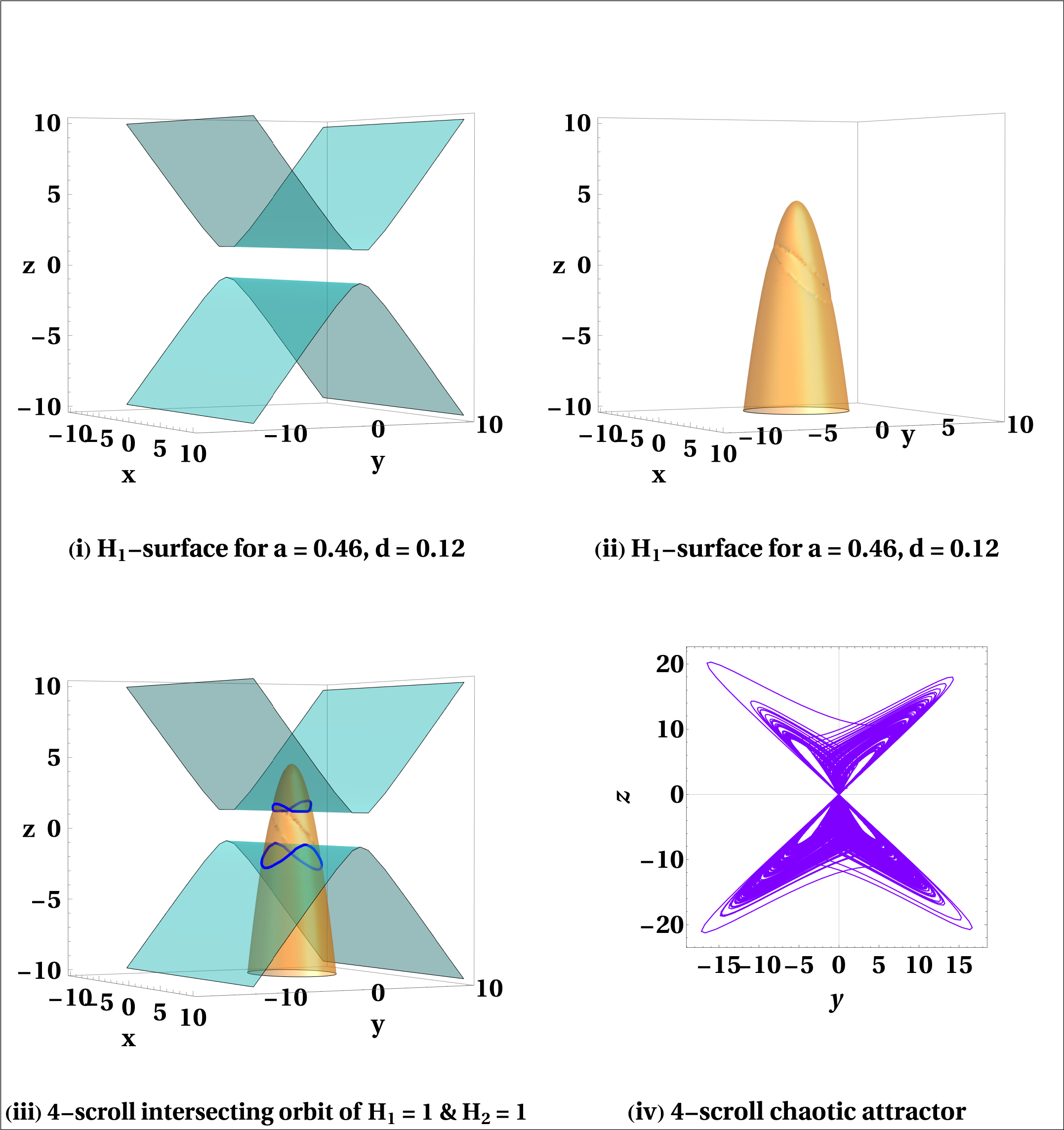}}
	
	\caption{Here, (a-d)(i) and (a-d)(ii) illustrate geometrical shape of $H_1$ and $H_2$ surfaces for different (a,d)-parameters. Again, (a-d)(iii) represent the intersecting orbit(shown in blue colour) of two constant Nambu surfaces shown in equation (\ref{eq20}) for different set of (a,d)-parameters, indicating the existence of 1,2,3,4-scrolls at some initial condition. And (a-d)(iv) represent the collection of large number of intersecting orbits at different instants, produced with the help of dissipative-Nambu mechanics. }

	\label{fig2}
\end{figure}

 In FIG. \ref{IOs}, we have illustrated a four set of parameter values, for which n-scroll chaotic attractors are produced in phase space as shown in sub-figures of FIG. \ref{IOs}(a). Here, 1,2,3,4-scroll chaotic attractors are represented in red, orange, green, sky blue colours and the corresponding equilibrium points of each attractor is denoted in blue dots. And we can clearly see that each scroll of 1,2,3,4-scroll attractor is associated with a particular equilibrium point. Again, these 1,2,3,4-scroll intersecting orbits can be emerged from the intersection of $H_{1}=k1$ \& $H_{2}=k2$ surfaces at certain initial condition and of course, both the surfaces are considered at the same initial condition. The sub-figures of FIG. \ref{IOs}(b) confirmed that we can generate all 1,2,3,4-scroll intersecting orbit for parameters ($a=1.48, d=0.16$),($a=3.5, d=0.06$),($a=0.977, d=0.1$),($a=1.46, d=0.06$) respectively. And the superposition of a large number of intersecting orbits at different time instants can produce corresponding full n-scroll attractors with the help of Nambu mechanics along with some dissipation involved and the geometrical interpretation of these n-scroll attractors are shown in sub-figures of \ref{IOs}(c). 
In FIG. \ref{IOs}, we have demonstrated the emergence of n-scroll attractors for different combination of (a,d)-values, where both a and d parameters have very small values. Again, in FIG. \ref{fig2}, we reproduce 1,2,3,4-scroll chaotic attractors using Nambu framework as a result of superposition or a collection of  several intersecting orbits of both Nambu surfaces $H_{1}=k1$ \& $H_{2}=k2$ for another four combinations of (a,d)-values and here, both a and d parameters have large values such as ($a=2, d=4.2$),($a=1.6, d=2.5$),($a=0.7, d=1$),($a=0.46, d=0.12$) respectively.

  We found that the intersecting orbits (shown in blue colour in FIG. \ref{fig2} (a-d)(iii)) are always closed homoclinic orbits and these are nothing but the non-dissipative phase-space trajectories of n-scroll chaotic attractors. A particular scroll/lobe is generated from the intersection of Nambu doublets at a particular instant, where both the doublet are considered at the same initial condition.The corresponding (k1,k2)-values for a certain initial condition are also mentioned. Hence, we conclude that when dissipation is not involved in the dynamics, $H_{1}$ and $H_{2}$ denotes two constants of motion. Similarly, by using the dissipative-Nambu mechanics(where dissipation part comes into account) shown in equation(\ref{equiv}), the full chaotic attractors are produced.

\begin{figure}[htbp!]
	{\includegraphics[width=0.87\linewidth]	{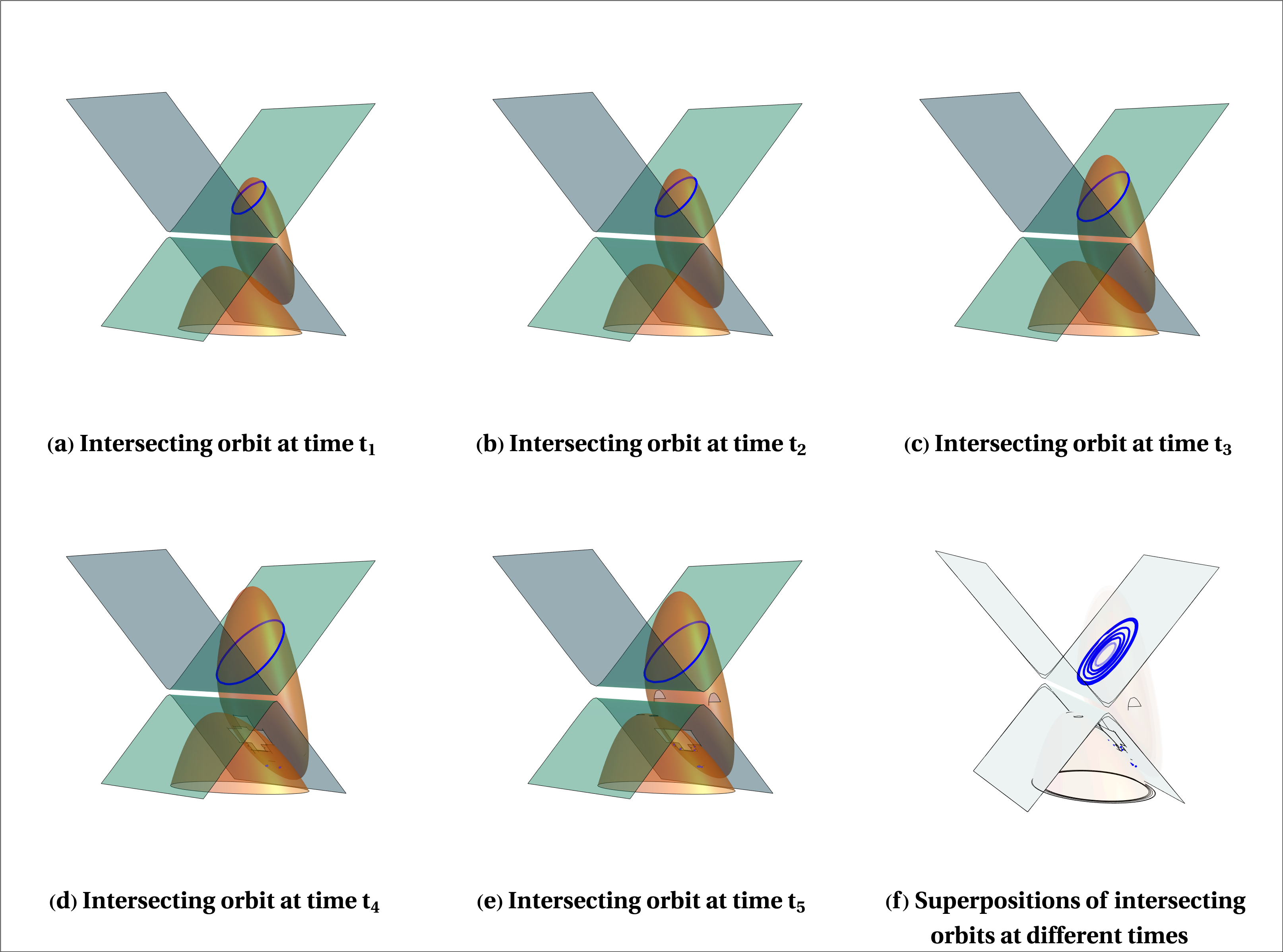}}
	
	\caption{The sub-figures (a-e) represent the snapshot of intersecting orbits of two Nambu-Hamiltonian surfaces $H_1$ and $H_2$  at some particular times $t_1$,$t_2$,$t_3$,$t_4$,$t_5$ ; indicating instantaneous trajectories of a 1-scroll chaotic attractor for control parameters $(a,d)$=$(2, 4.2)$. And the sub-figure (f) illustrates the discrete time evolution of a 1-scroll chaotic attractor trajectories, which is represented by the superpositions of several intersecting orbits at various distinct times. }
	\label{collc1}
\end{figure}
\begin{figure}[htbp!]
	{\includegraphics[width=0.87\linewidth]	{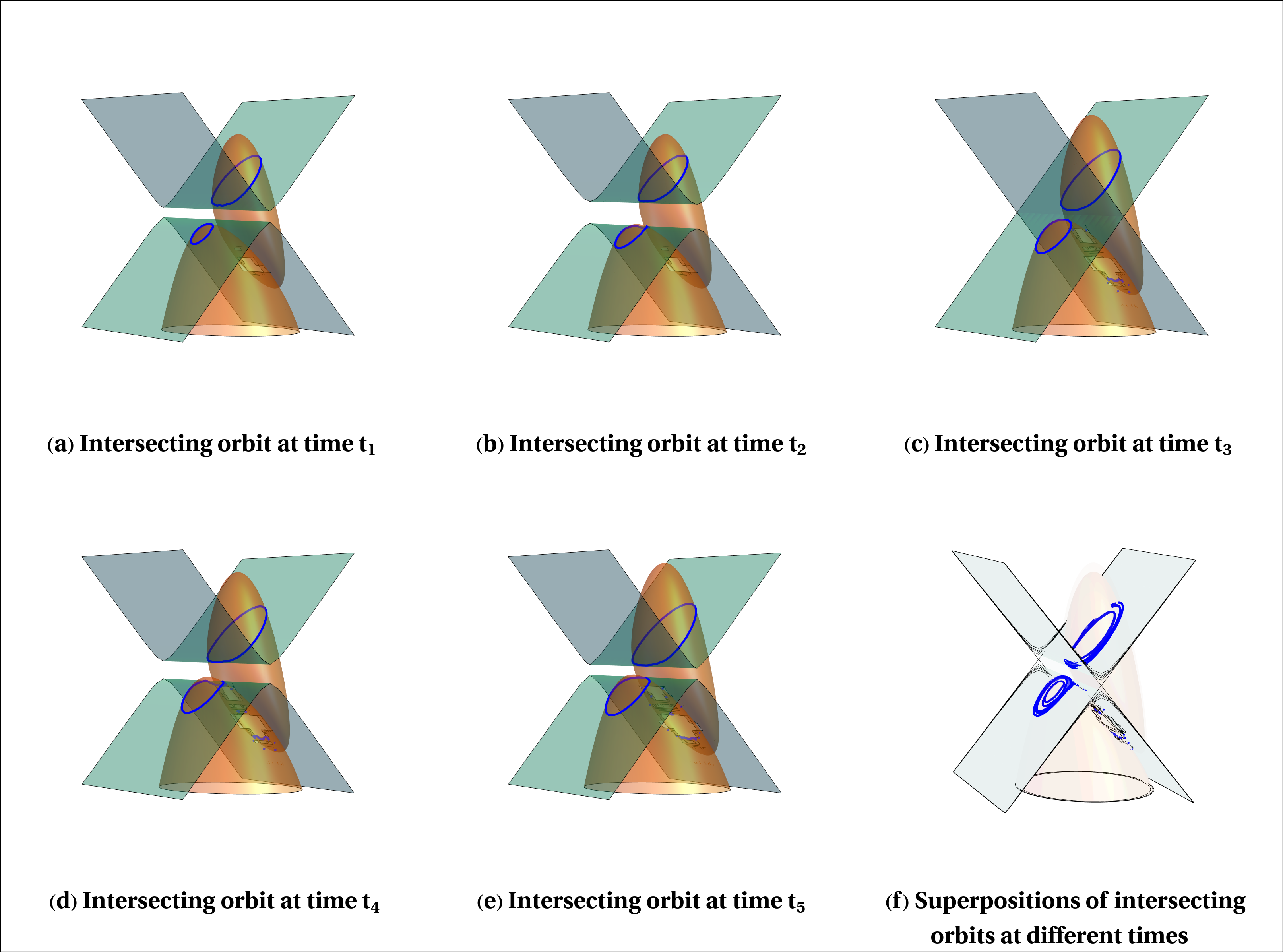}}
	
		\caption{The sub-figures (a-e) represent the snapshot of intersecting orbits of two Nambu-Hamiltonian surfaces $H_1$ and $H_2$  at some particular times $t_1$,$t_2$,$t_3$,$t_4$,$t_5$ ; indicating instantaneous trajectories of a 2-scroll chaotic attractor for control parameters $(a,d)$=$(1.6, 2.5)$. And the sub-figure (f) illustrates the discrete time evolution of a 2-scroll chaotic attractor trajectories, which is represented by the superpositions of several intersecting orbits at various distinct times. }
	\label{collc2}
\end{figure}
\begin{figure}[htbp!]
	{\includegraphics[width=0.87\linewidth]	{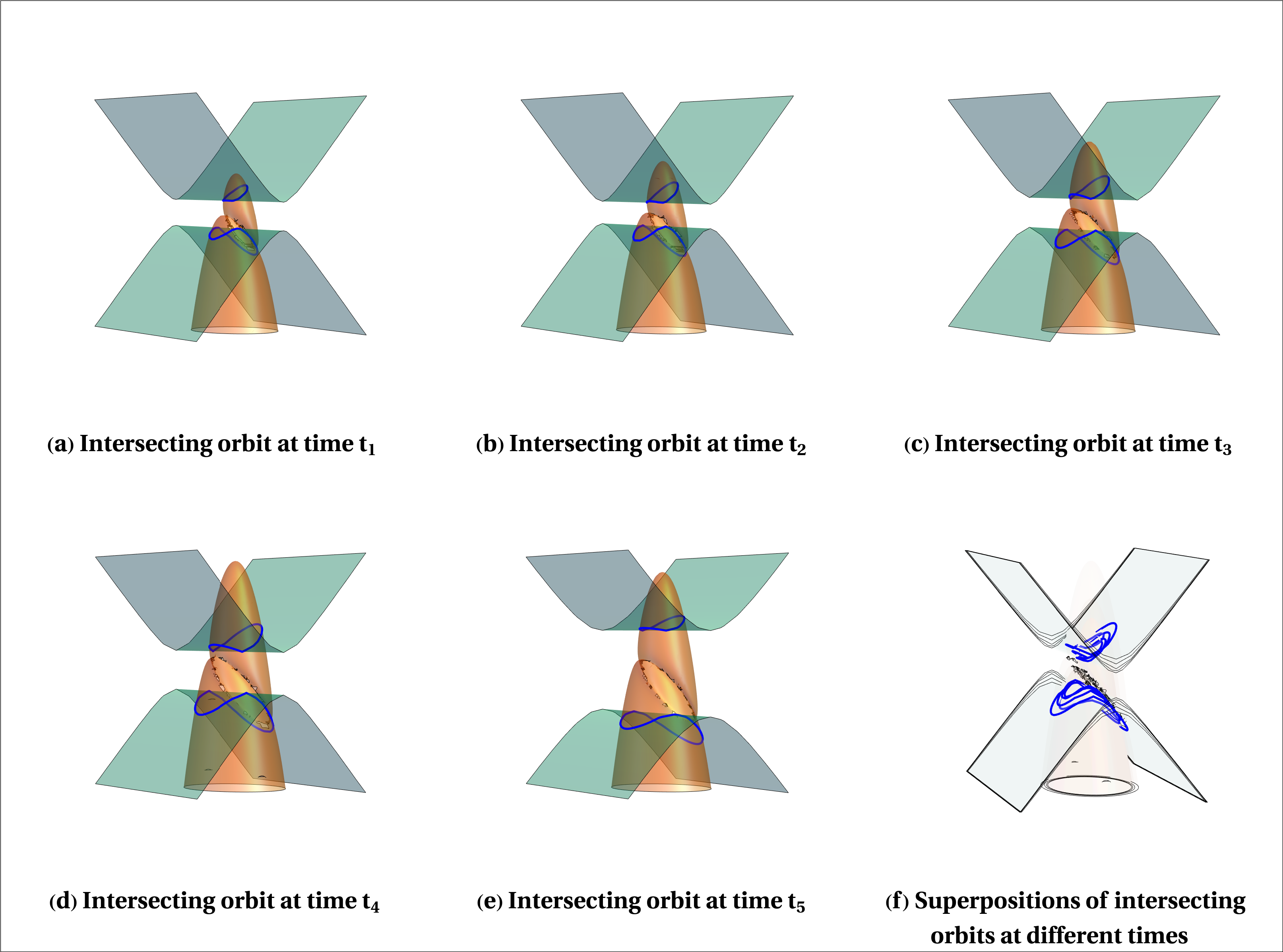}}
	
	\caption{The sub-figures (a-e) represent the snapshot of intersecting orbits of two Nambu-Hamiltonian surfaces $H_1$ and $H_2$  at some particular times $t_1$,$t_2$,$t_3$,$t_4$,$t_5$ ; indicating instantaneous trajectories of a 3-scroll chaotic attractor for control parameters $(a,d)$=$(0.7, 1)$. And the sub-figure (f) illustrates the discrete time evolution of a 3-scroll chaotic attractor trajectories, which is represented by the superpositions of several intersecting orbits at various distinct times. }
	\label{collc3}
\end{figure}
\begin{figure}[htbp!]
	{\includegraphics[width=0.87\linewidth]	{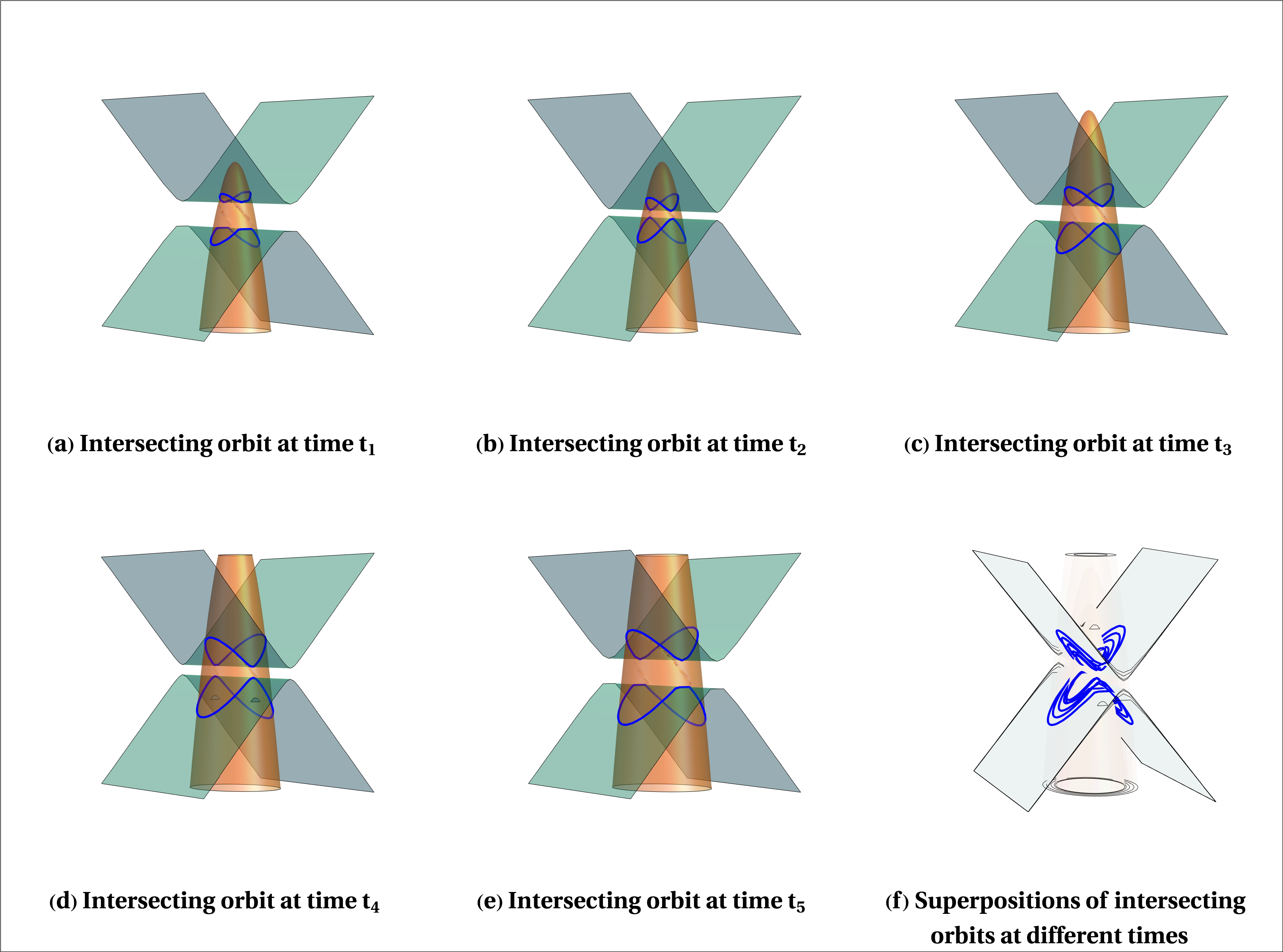}}
	
	\caption{The sub-figures (a-e) represent the snapshot of intersecting orbits of two Nambu-Hamiltonian surfaces $H_1$ and $H_2$  at some particular times $t_1$,$t_2$,$t_3$,$t_4$,$t_5$ ; indicating instantaneous trajectories of a 4-scroll chaotic attractor for control parameters $(a,d)$=$(0.46, 0.12)$. And the sub-figure (f) illustrates the discrete time evolution of a 4-scroll chaotic attractor trajectories, which is represented by the superpositions of several intersecting orbits at various distinct times. }
	\label{collc4}
\end{figure}
FIG. \ref{fig2}  illustrates how different number of wings geometry may form depending on the parameters values. The key reason for the formation of desired intersecting loops depends on two factors. First the specific geometry of the surfaces, second relative orientation of the two surfaces. In all the cases one of the surface remains hyperbolic two sheets surfaces for all values of the parameters. Each surface has zero slope parallel to the X-axis, as if two sub-surfaces meet at a line parallel to X-axis. Therefore we can imagine the two sheets hyperbola consisting of four sub-surfaces. Because of this geometry, four separate loops may form. The other surface is ellipsoid whose deformation depends on the system parameter as shown in the FIG. \ref{fig2}. A single loop is formed when the ellipsoid intersect with only one hyperbolic sub-surface as the case in FIG. \ref{fig2}(a). The four separate loops are possible when the deformed ellipsoid intersects with all the hyperboloid surface, as shown in the FIG. \ref{fig2}(d), giving rise to 4-wing chaotic dynamics. This result clearly shows the origin of formation of multi wing due to multi-scroll dynamics lies with the geometry of the Nambu surfaces.\\

When dissipative part is taken into consideration, the full system does not conserve $H_{1}$, $H_{2}$ and the Nambu surfaces become time dependent functions. This results in continuous change of relative orientation and area of the two surfaces. Therefore, the intersecting orbits of these surfaces-which as a whole represent the trajectories of the full system-does also change with time. Hence, this dynamical pictures shown in FIG. \ref{fig2}(a-d)(iv) can also be interpreted as the collection(or superpositions) of large number of intersecting orbits, generated by the equation(\ref{sur1}), at different instants.
FIG. \ref{collc1}-\ref{collc4}  illustrate the transitions of intersecting orbits of corresponding Nambu surfaces at different times followed by the formation of the full n-scroll or multi-lobe chaotic attractors. Hence, we can say that the visual representation of multi-lobe chaotic structures is powerfully enhanced through the application of Nambu mechanics, offering a compelling perspective on their complexity.



\section{\label{secV}Conclusion}
The dynamical system under study has several equilibrium points, depending on the values of the system parameters. The stability of these equilibrium points are studied by linearization method and found to have both stable and unstable equilibrium points. By numerical solution, we plotted phase space of the system and found distinct dynamical scrolls, originating from  the calculated equilibrium points. To investigate these 3-D autonomous systems through the framework of Nambu mechanics, we employ a systematic decomposition into two fundamental components: an irrotational non-dissipative subsystem and a rotational dissipative subsystem. We observe that the geometry of Nambu surfaces for different system parameters at which different scrolls appear have distinctly different geometries. These different geometries are responsible for desired intersection of the two surfaces which give rise to desired wings like appearance in the phase space. This fundamental result show that the origin of wing like structure is specific geometry of Nambu surfaces which can be demonstrated without numerically solving the dynamical equations. These different wings due to intersection of two surfaces found to swirl about the equilibrium points, thus verify the numerical result. We have also demonstrated how the full dynamics generate when the dissipative part is added to the non-dissipative part.




\bibliography{References}

\end{document}